\documentclass[3p,twocolumn]{elsarticle}

\usepackage[mathdef]{custom} 

\crefname{equation}{}{}

\newcommand{\norms}[1]{\lVert#1\rVert}

\newcommand{\normb}[1]{\norm{\bs{#1}}}
\def\tq{\tilde{\bs{q}}}
\def\hq{\hat{\bs{q}}}
\newcommand{\pde}[2]{\frac{\partial #1}{\partial #2}}

\makeatletter
\newcommand{\checkl}[2]{
  \@ifundefined{r@#1}{}{#2}
  }
\makeatother

\tikzstyle{basic box}[red] = [shape=rectangle, align=center, draw=#1, rounded corners, text width=90pt]
\tikzstyle{basic box b}[red] = [shape=rectangle, align=center, draw=#1, rounded corners]
\tikzstyle{state} = [circle, draw, very thick]

\journal{International Journal of Electrical Power and Energy Systems}

\begin{document}

\begin{frontmatter}

\title{ 
Incentive Mechanisms to Prevent Efficiency Loss of Non-Profit Utilities
}

\author[carlos]{Carlos Barreto\corref{correspondingauthor}}
\address[carlos]{Department of Electrical Engineering and Computer Science, Vanderbilt University, Nashville, TN, United States}
\cortext[correspondingauthor]{Corresponding author}
\ead{carlos.a.barreto@vanderbilt.edu}

\author[eduardo]{Eduardo Mojica-Nava}
\address[eduardo]{Departamento de Ingeniería Eléctrica y Electrónica, Universidad Nacional de Colombia, Bogotá, Colombia}
\ead{eamojican@unal.edu.co}

\author[nicanor1]{Nicanor Quijano} 
\address[nicanor1]{Departamento de Ingeniería Eléctrica y Electrónica, Universidad de los Andes, Bogotá, Colombia}
\ead{nquijano@uniandes.edu.co}

\begin{abstract}
\modification{

The modernization of the power system introduces technologies that may improve the system's efficiency by enhancing the capabilities of users. 
Despite their potential benefits, such technologies can have a negative impact. 
This subject has widely analyzed, mostly considering for-profit electric utilities.
However, the literature has a gap regarding the impact of new technologies on non-profit utilities.

In this work, we quantify the price of anarchy of non-profit utilities, that is, the cost caused by lack of coordination of users.
We find that users, in the worst case, can consume up to twice the optimal demand, obtaining a small fraction of the optimal surplus.
For this reason, we leverage the theory of mechanism design to design an
incentive scheme that reduces the inefficiencies of the system, which preserves the privacy of users.
We illustrate with simulations the efficiency loss of the system and show two instances of incentive mechanism that satisfy either budget balance and budget deficit.
}

\end{abstract}


\end{frontmatter}

\section{Introduction}

\modification{
Power systems are experiencing a innovation process to improve multiple aspects, such as efficiency, reliability, and security. 
Although new technologies 
play an important role,  the current modernization endeavor requires the active participation of customers.
In particular, \emph{demand response} programs leverage communication and automation technologies so that customers can manage their loads according to the system's state.
In this way, customers can make better use of the available resources and offer services, such as  help
reducing the stress on the system or sell energy.

The literature on demand response focus mainly on how
to achieve different goals, such as reduce demand peaks and/or improve the efficiency of the system \cite{vardakas2015survey, aliabadi2017competition}. 
%
%
These research works often consider investor owned utilities, that is, firms focused on maximizing their profits;
however, little attention has been given to non-profit electric utilities (e.g., regulated or customer owned cooperatives).

Although investor owned utilities dominate the electricity markets, non-profit electric cooperatives play an important role in the economic development \cite{paredes2018rural}.  
Besides,  electric cooperatives have properties that foster the social support for renewable energy projects \cite{heras2018emergence, capellan2018renewable, huybrechts2014relevance}.\footnote{The  REScoop.eu network has 1500 renewable energy cooperatives in Europe \cite{rescoop}.}
On one hand, cooperatives offer transparency in the prices and allow customers to control the source of energy.
Thus, the owners/members of an electric cooperative can choose to support environmental or social objectives,
 rather than maximize profits, as investor owned firms.

On the other hand, cooperatives present the renewable energy infrastructures as community owned, which promotes their acceptance.
Also, cooperatives can inform about environmental issues and give advice on how to reduce the environmental impact, e.g., reducing the consumption of energy.\footnote{Other electric corporations profit with sales, so they do not have incentives to promote reductions in demand.}
According to \cite{huybrechts2014relevance}, the customers of Ecopower, a renewable energy cooperative from Belgium, reduced their average demand from 4000 kWh/year to 3000 kWh/year within 3 years.

In this work we close a gap in the literature 
designing a demand response approach for non-profit utilities.\footnote{Here we refer to non-profit utilities as customer owned utilities, regulate utilities, or electric cooperatives.}  
First, we show that behavioral changes enabled with new technologies can harm the system's efficiency to a large degree.
This motivates the design of an 
incentive scheme that reduces efficiency losses. 
This research is important for electric utilities, which face the challenge of adapting their policies to account for new capabilities of their customers.

\subsection{Literature Review}

The demand response literature often considers the following objectives:
reduce or shift the demand peak, maximize the \emph{social welfare} (the  social satisfaction with an allocation of resources), minimize costs, or some combination of these goals \cite{vardakas2015survey, deng2015survey}.
Maximizing the social welfare may encompass the other objectives, because  
1) maximizing the social welfare, assuming inelastic demand, equals to minimizing costs, and 2) 
reduction of demand peaks may be necessary to improve the social welfare.
However, these goals do not represent the objective of regulated or non-profit utilities, because they prioritize the customer's satisfaction over profits.

In general, both non-profit and regulated utilities may choose electricity tariffs so that their revenues  amounts to the costs incurred to provide the service.
The cost aggregates both the operating expenses (e.g., fuel, labor, and maintenance) 
and the 
cost of the capital invested (e.g., interest, debt, and a fair rate return to investors) \cite{rap2016revenue}.
In particular, the \emph{average cost pricing} scheme guarantees that the utility collects the estimated total cost.

Game theory has been widely used in demand response studies to analyze interactions of selfish agents with conflicting interests   \cite{mohsenian, CLLD10, li2016market, chen2014autonomous, corchon1994comparative, de2017convergence}.
In this case, the conflict arises because customers may compete for a limited resource.
Some research on demand response focus on setting up the conditions (e.g., prices, incentives) so that selfish agents reach a desired social goal. 
In other words, demand response attempts to coordinate agents whose individual actions could harm the society.
%
%

The seminal work in \cite{koutsoupias1999worst}
introduces the \emph{price of anarchy}, a measure to estimate the impact of lack of coordination of agents.
In our context, the price of anarchy can quantify the effect of new technologies that, although enhance the decision making capabilities of customers, may not improve their cooperation.
Thus, we can think of demand response programs as efforts to reduce the price of anarchy (i.e., to improve the efficiency of the system).

Previous works analyze possible problems created by new technologies, such as stability \cite{zhou2017stability, roozbehani2012volatility} or
security issues \cite{barreto2018impact, blackiot, gross09, liu2014vulnerability, LiyanThomasTong12}. In a way, these works analyze the cost of ill designed demand response schemes.
However, the price of anarchy offers a different perspective, which measure the cost of not having demand response at all.

Some efforts have been made to analyzed the price of anarchy of different systems.
For example, 
\cite{tardos} analyzes the degradation of a network's performance due to unregulated traffic.
Moreover, 
\cite{johari2004efficiency} studies how users who anticipate the effect of their actions reduce their utility in congestion games.
However, we are not aware of works that analyze the price of anarchy of power systems with non-profit utilities.


The literature has several mechanisms to prevent efficiency losses in problems of resource allocation.
%
%
For example, 
\cite{johari2006scalable} proposes a resource allocation mechanism 
 that has a bounded efficiency loss when users anticipate prices.
On the other hand, 
\cite{samadi2012advanced} 
presents a
 pricing mechanism to implement efficient outcomes in smart grids.
Although relevant, these works focus on maximizing the social welfare, that is, they consider for-profit firms.
Moreover, \cite{sandholm2005negative} proposes the evolutionary implementation of a mechanism that guarantees efficiency and stability. 
This mechanism assumes that the payoff functions are additively separable in the private information of users.

Works such as \cite{teneketzis12a, teneketzis12b} propose decentralized mechanisms to allocate efficiently bandwidth in networks. These mechanisms satisfy the \emph{budget balance} property,\footnote{The mechanism does not need external subsidies nor imposes taxes on the users.} but require a multidimensional message space, because users must report their desired allocation and the price they are willing to pay for each link of the network.

Previous works analyze the dependence of the message space with the efficiency of the mechanism. 
For instance, \cite{reichelstein1988game} analyzes the minimum space required to implement optimal allocations in exchange economies, which is larger than the number of participants and resources. 
Furthermore, 
\cite{healy2012designing} shows that one-dimensional message spaces are unsuitable to achieve optimal allocations on contractive games (games whose best response is a contraction mapping).

\subsection{Contributions}

At a high level, we make the following contributions:
\begin{itemize}
\item We analyze the efficiency loss (price of anarchy) in power systems with non-profit utilities (most of the literature in demand response considers for profit-utilities). 

\item We propose an incentive mechanism to prevent inefficiencies and preserves the privacy of the customers.  

\item We illustrate the efficiency loss and the  efficacy of the mechanism through simulations. 
\end{itemize}

We leverage the theory of mechanism design to propose an incentive scheme that reduces the efficiency loss. 
Our incentives scheme is based on the Clarke pivot mechanism \cite{AlgorithmicG} and 
preserves private information using
a one-dimensional message space per each resource to be allocated.\footnote{A one-dimensional message space reduces the technical requirements of the system and also protects the privacy of the customers, since they do not report their electricity consumption preferences.
}
We manage to design a mechanism that uses a one-dimensional space (to allocate a single resource) thanks to the properties of the system. 
This is possible because the wealth of each agent depends on their actions and the aggregate actions of other agents. Thus, we can decentralize decisions by sending to each agent the aggregated demand, which is one-dimensional. Likewise, users report their demand (a one dimensional signal) to the utility.
The structure of the proposed mechanism is similar in philosophy to the mechanism in \cite{Groves&Ledyard1977}; however, 
in our case users do not communicate their marginal payoff function.

We extend the works  previously published in \cite{barreto2013design, barreto2015incentives} in several ways: 
1) we find that the efficiency loss takes place with an overuse of resources (in the inefficient equilibrium the demand can double the optimal value);
2) we prove that, in the worst case, the efficiency loss with strategic agents is arbitrarily large;
3) we show two instances of incentives that satisfy individual rationality (i.e., users have always positive surplus) and have either budget deficit or budget surplus;
and 4) we generalize previous results relaxing restrictions on the price function, we only assume that the generation cost function is strictly convex.

\subsection{Organization of the Paper}

In \cref{sec:backgnd} we introduce the electricity system model.
We show that the average pricing scheme leads to optimal outcomes when users have limited capabilities, i.e., when they lack technologies to plan strategically their actions. 
Then, we show that strategic behaviors harm the systems efficiency. In particular, the main result of the paper shows that users have incentives to consume up to twice the optimal demand, which in turn reduces significantly the system's efficiency. 
In \cref{sec:incentives} we propose and analyze an incentives scheme to reach the optimal equilibrium with strategic users. 
We show two instances of the incentives that satisfy different properties.
Finally, we conclude the paper in \cref{sec:concl}.

}



\section{Background}\label{sec:backgnd}

\modification{
In this section we introduce a model of the power system\footnote{Here we assume that a non-profit electric utility provides both generation, transmission, and distribution. Thus, the electric utility manages the whole system.} 
and analyze the system's equilibrium with two types of users, namely non-strategic and strategic, which describe users before and after the introduction of DR technologies, respectively.
The main results of this section show that strategic behaviors damage the efficiency of the electricity system. 
In the worst case, strategic customers consume twice the optimal amount of energy, causing an arbitrarily large  efficiency degradation.
}

\modification{
\subsection{Notation}
We denote vectors with \textbf{bold} font. Let $\bf{q}$ be a vector of size $N$. Thus, we define $\boldsymbol{q}_{-i}$ as a vector with the elements of  $\boldsymbol{q}$, except the $i^{th}$, i.e., $\boldsymbol{q}_{-i} = [q_1, \ldots, q_{i-1}, q_{i+1}, \ldots, q_N]$. 
The norm $\left\Vert \cdot \right\Vert$ represents the $L^1$-norm, that is,  $\left\Vert \bf{q} \right\Vert = \sum_{i=1}^N |q_i|$. 
Finally,  $\dot f$ and $\ddot f$ denote the first and second derivatives of the function $f$, respectively.
\cref{tab:nomenclature} summarizes the symbols used in the document.

\begin{table}[tb]
\caption{List of symbols.}
\label{tab:nomenclature}
\centering
\begin{tabular}{cll} \toprule
Symbol & & Description \\ \midrule
$N$ & & Number of customers \\
$\mathcal{P}$ & & Set of customers \\
$q_i$ & & Demand of the $i\th$ customer \\
$\boldsymbol{q}$ & & 
Demand of the population \\
$g$ & & Total demand \\
$C$ & & Generation cost function \\
$p$ & & Unitary price of energy \\
$t_i$ & & Payment of the $i\th$ customer \\
$v_i$ & & Valuation function \\
$U_i$ & & Profit of the $i\th$ customer \\
$\sigma_i$ & & Set of available strategies \\
$\mathcal{G}$ & & Game \\
$\boldsymbol{\mu}$ & & Optimal demand \\
$\boldsymbol{\xi}$ & & Nash equilibrium \\
$r$ & & Efficiency ratio \\
$\mathcal{M}$ & & Mechanism \\
$I$ & & Incentive function \\
$I^d$ & & Incentive with budget deficit \\
$I^s$ & & Incentive with budget surplus \\
$W_i$ & & Profit 
with the incentives \\
$\mathcal{G}_\mathcal{I}$ & & Game with incentives \\
$h$ & & Estimation of the electricity price \\
$o$ & & Outcome function \\
\bottomrule
\end{tabular}
\end{table}

}

\subsection{Model of the Electricity System}\label{sec:model}

We consider a 
non-profit
utility that provides energy to a population $\mathcal{P} = \{1,\ldots,N\}$ with $N$ customers.
%
Let us denote by  $C(g)$ the total cost to supply a demand of  $g$ units of energy.\footnote{$C(g)$ includes operation costs, such as generation, transmission, and a fair return for the utility.}
For simplicity, we ignore transmission losses, therefore, the total energy generated equals the total demand,
i.e.,  
$g = \sum_{i\in\mathcal{P}} q_i$ , where $q_i\geq 0$ is the demand of the $i^{th}$ user.

We define the profit of users as  the benefit earned minus the cost of the energy consumed.
Here we quantify the benefit of the $i^{th}$ user through a  \emph{valuation function} $v_i:\mathbb{R}_{\geq0}\rightarrow \mathbb{R}$, where $v_i(q_i)$ represents the monetary value that the user assigns to $q_i$ units of energy.\footnote{The valuation function reflects the benefit consuming $q_i$ units of energy. \modification{
For instance, a concave valuation function, such as 
 $v_i (q_i) = \log(1+q_i)$, indicates that the value of an additional resource decreases with the total number of resources $q_i$.}
}
Besides, the $i^{th}$ user pays $q_i p(g)$ for its consumption, where $p(g)$ is the unitary price of energy, which depends on the total demand $g$.
In summary,  the profit of the $i^{th}$ consumer is 
\begin{equation}\label{eq:utility}
U_i(q_i, g) = v_i(q_i) -  q_i p(g).
\end{equation}

\modification{
We make the following assumption regarding the functions of the model.
\begin{assumptionletter} \label{as:1} \
%
%
\begin{enumerate}

 \item The valuation function $v_i:\mathbb{R}_{\geq0}\rightarrow \mathbb{R}$ is twice differentiable, concave,  non-decreasing, and satisfies $v_i(0) = 0$.

 \item The cost function  $C:\mathbb{R}_{\geq 0} \rightarrow \mathbb{R}_{\geq 0}$
 is differentiable, strictly convex, and satisfies $C(g) > 0$ if $g>0$.

 \item The average cost $p(g) = \frac{C(g)}{g}$ is monotonically increasing, therefore, $\dot{p} (g) \geq 0 $ for $g > 0$.
\end{enumerate}
 %
 
\end{assumptionletter}
}

The electricity tariff $p(\cdot)$ and the
behavior of users determine the equilibrium of the system.\footnote{The equilibrium of a power system has the form of a \emph{Nash equilibrium}, which describes situations where no user has incentives to change its demand unilaterally (see Definition \ref{def:nash}).} 
In traditional power systems users cannot observe changes in prices, therefore, we assume that they 
cannot manifest strategic behaviors.
However, new technologies grant users more information and capabilities to make decisions. Hence, users can 
develop strategies 
to anticipate the effect of their actions in the prices \cite{AlgorithmicG}. 
Below we 
show that the average cost price is the optimal tariff 
in systems with 
non-strategic users. 
Latter we prove that the 
self-interest of strategic 
users can degrade the efficiency of 
the \modification{electricity system}. 

\subsection{System with Non-strategic Users}

Here we define the behavior of users before introducing DR technologies.
Let us denote the surplus of non-strategic users as 
\begin{equation}\label{eq:surplus_price_takers}
 \tilde{U}_i(q_i, \tilde{p}) = v_i(q_i) - q_i \tilde{p} ,  \quad i \in \mathcal{P},
\end{equation}
where $\tilde{p}\in \mathbb{R}$ is the price observed by the users.
Rational users will select the demand that maximizes their surplus, given the constant price $\tilde{p}$ that they observe.
 Hence, each user solves the following optimization problem
\begin{equation}
  \begin{aligned}
  \underset{q_i}{\text{maximize}}
  & \quad \tilde{U}_i(q_i, \tilde{p}) = v_i(q_i) - q_i \tilde{p}  \\
  \text{subject to}
  & \quad q_i \geq 0. \\
  \end{aligned}
\end{equation}
From  \cref{as:1} the surplus in \cref{eq:surplus_price_takers} is concave; hence, the previous problem has a unique solution, denoted by $\tilde{q}_i$, which satisfies \cite{boyd2004convex, kuhn1951proceedings, karush1939minima}
\begin{equation}\label{eq:foc_price_takers}
\begin{cases}
\tilde{q}_i = 0 & \quad \text{if } \dot v_i(0) < \tilde{p},  \\
 \dot{v}_i (\tilde{q}_i) - \tilde{p} = 0 &  \quad \text{otherwise}.
\end{cases}
\end{equation}
The previous expression implies that all users with positive demand ($\tilde{q}_i > 0$) have the same marginal valuation in the equilibrium.

\modification{The utility
can select price schemes} that allow users 
reach efficient demand profiles (or allocations of resources) acting independently.
Below, we 
introduce the concept of \emph{Pareto optimal allocation} and 
leverage the \emph{second welfare theorem} to show that the \emph{average pricing} scheme leads to a Pareto optimal allocation.\footnote{In a Pareto optimal allocation no user can be made better off without another being made worse off.}

\subsubsection{Pareto Optimality}

We define the Pareto efficiency criteria as follows  \cite{mas1995microeconomic}:
\begin{definition}
 An allocation $\tilde{\bs{q}}=[\tilde{q}_1, \ldots, \tilde{q}_N]$ is Pareto optimal (or Pareto efficient), given some price $\tilde{p}$, if there is no other allocation $\hat{ \bs{q} }$ such that $\tilde{U}_i(\hat{q}_i, \tilde{p}) \geq \tilde{U}_i(\tilde{q}_i, \tilde{p})$ for all $i\in \mathcal{P}$, with strict inequality for some $i\in\mathcal{P}$.
\end{definition}
From this definition, it follows that an allocation $\tq$ that maximizes the \emph{consumer surplus}, defined as $\sum_i \tilde{U}_i(\tilde{q}_i, \tilde{p})$, is Pareto optimal.
In other words, no other allocation $\hat{\bs{q}}$ would improve the benefit of users without affecting any of them, otherwise, $\hat{\bs{q}}$  would attain a higher social surplus than $\tq$, which leads to a contradiction.

\subsubsection{Optimality of the Average Cost Pricing}
With the following optimization problem we express the problem of finding
the system's equilibrium, represented by the tuple $(\tilde{p}, \tq)$,
that maximizes the customer surplus
%
\begin{subequations}
  \begin{align}
  \underset{ q_1,\ldots, q_N,  p }{\text{maximize}}
  & \quad \sum\nolimits_{i\in\mathcal{P}} \tilde{U}_i (q_i, p) 
  \label{eq:agg_surplus_price_takers}  \\
  \text{subject to}
  &  \quad C\left(\normb{q}\right) - \normb{q} p \leq  0, \label{eq:cost_res} \\
  & \quad q_i (\dot{v}_i(q_i) - p) = 0 , \quad i\in\mathcal{P} \ \label{eq:users_res} \\
  & \quad q_i\geq 0, \quad i\in\mathcal{P}.
  \end{align}
  \label{eq:opt_price_takers}
\end{subequations}
The restriction in 
\cref{eq:cost_res} is a budget constraint required to guarantee that the total payments by the users $\left( \normb{q} p \right)$ cover the production costs $C\left( \normb{q} \right)$.\footnote{In practice, the operational restrictions of some generators (e.g., large start up costs) may force them to run even when they incur in losses, provided that they compensate the losses in future periods. Hence, the restriction in \cref{eq:cost_res} refers to the long run operation \cite{kirschen2004fundamentals}.}
Moreover,  
\cref{eq:users_res}
captures the optimal response of the users (conditions in \cref{eq:foc_price_takers}). 
The following result shows that a system with the average cost pricing has a Pareto efficient equilibrium that maximizes the customer surplus.
\begin{theorem}\label{thm:avg_prices}
Consider non-strategic users with surplus given by \cref{eq:surplus_price_takers} and a generator with cost $C(\cdot)$ satisfying Assumption \ref{as:1}. Then, the average cost price function
\begin{equation}\label{eq:avg_cost}
 p\left( g \right) = \slfrac{C\left( g \right)}{g},
\end{equation}
leads to a Pareto equilibrium.
\end{theorem}
The Appendix contains the proof of this and the remaining results.

\subsection{System with Strategic Users}

In this subsection we formulate the profit of strategic users as
 a function of both the total demand 
 $g=\normb{q}$ 
 and the user's demand $q_i$\footnote{
Users must reveal their consumption only to the utility company, which in turn computes and broadcast the aggregate demand.
}
\begin{equation}\label{eq:profit_strategic}
 U_i \left(q_i, \bs{q}_{-i} \right) = v_i(q_i) - t_i(\bs{q}),
 \quad i\in\mathcal{P},
\end{equation}
where 
\begin{equation}\label{eq:t_i}
t_i(\bs{q}) = q_i p \left( \norm{\bs{q}} \right)
\end{equation}
represents the payment function of the $i^{th}$ user.
We make the following assumption to guarantee that the
profit of the individuals $U_i(q_i, \bs{q}_{-i})$ is concave.
\begin{assumptionletter}\label{as:2}
 The payment $t_i(\bs{q})$ in \cref{eq:t_i} is increasing convex with respect to $q_i$.
 Furthermore, the marginal  payment $\pde{t_i(\bs{q})}{q_i} = p\left(\normb{q}\right) + q_i \dot p\left(\normb{q}\right)$  is increasing with respect to $q_i$. 
\end{assumptionletter}

As users improve their decision capabilities, 
the system's equilibrium changes; however, the system's efficiency may decrease. 
\modification{
Before analyzing the efficiency of the system, we need introduce 
both the optimal equilibrium and 
the equilibrium  that users reach pursuing their own interests.
%
Later we investigate the relation between these equilibria.
}

\subsubsection{Optimal Equilibrium}
\label{sec:optimal_equilibrium}

The demand $\bs{\mu}\in\mathbb{R}^N_{\geq 0}$ that maximizes the customer surplus (our efficiency metric), given the average cost price, solves the following optimization problem
\begin{equation}\label{eq:max_individual_profit_a}
  \begin{aligned}
  \underset{q_1, \ldots, q_N}{\text{maximize}}
  & \quad \sum_{i\in\mathcal{P}} U_i \left(q_i,  \bs{q}_{-i} \right) = \sum_{i\in\mathcal{P}} v_i(q_i) - C\left(\normb{q}\right) \\
  \text{subject to}
  & \quad q_i \geq 0, \quad i\in\mathcal{P}. \\
  \end{aligned}
\end{equation}
From \cref{as:1} the consumer surplus is strictly concave, therefore,  
there is a unique optimal solution
that satisfies the following conditions \cite{boyd2004convex}
\begin{equation}\label{eq:foc_optimal}
\begin{cases}
\mu_i = 0  & \quad \text{if } \dot{v}_i(0) < \dot{C}(\normb{\mu}),  \\
\dot{v}_i(\mu_i) = \dot{C}(\normb{\mu}) & \quad \text{otherwise}.
\end{cases}
\end{equation}

\subsubsection{Nash Equilibrium}
\label{sec:game}

We analyze the interactions of users using 
%
the  \emph{Cournot competition} model, in which users individually and simultaneously select the quantity that they want to consume \cite{cournot1838recherches, mas1995microeconomic}.
Particularly, we define the game as a 3-tuple  $\mathcal{G} = \langle \mathcal{P}, (\Sigma_i)_{i \in \mathcal{P}}, (U_i)_{i \in \mathcal{P}} \rangle$, where $\mathcal{P}$ is the set of players (or customers), $\Sigma_i = \mathbb{R}_{\geq 0}$ is the set of available strategies (consumption of electricity)
and $U_i: \Sigma_1 \times \dots \times \Sigma_N \to \mathbb{R}$ is the surplus function of the $i^{th}$ player. 
%

The equilibrium concept used in game theory is   the \emph{Nash equilibrium} (NE) \cite{nash1951non, fudenberg-tirole}, defined as follows. 
\begin{definition}\label{def:nash}
 The Nash  equilibrium of a game $\mathcal{G}$, denoted by $\bs{\xi} \in\mathbb{R}_{\geq 0}^{N}$, satisfies
$
 U_i (\xi_i,\bs{\xi}_{-i}) \geq U_i (q_i,\bs{\xi}_{-i})  , \, \text{for all } q_i\in\mathbb{R}_{\geq 0}, 
$
for every user $i\in\mathcal{P}$.
\end{definition}
From this definition we know that rational agents will choose $\xi_i$ when the other users choose $\bs{\xi}_{-i}$. In other words, $\xi_i$ is the demand that solves the following optimization problem
\begin{equation}\label{eq:max_individual_profit_b}
  \begin{aligned}
  \underset{q_i}{\text{maximize}}
  & \quad U_i \left(q_i,  \bs{\xi_{-i}} \right) \\
  \text{subject to}
  & \quad q_i \geq 0. \\
  \end{aligned}
\end{equation}
From \cref{as:2} we know that the problem in \cref{eq:max_individual_profit_b} is concave, and therefore, it has only one solution (given some $\bs{\xi}_{-i}$) that satisfies the following conditions
\begin{equation}\label{eq:foc_strategic}
\begin{cases}
\xi_i = 0 
  & \quad \text{if }  \dot{v}_i (0) <  p \left( || \bs{\xi} || \right), \\
 \dot{v}_i (\xi_i) = p \left( || \bs{\xi} || \right) + \xi_i \dot{p}\left( || \bs{\xi} || \right) & \quad \text{otherwise}.
\end{cases} 
\end{equation}
Moreover, we can use the results by Rosen \cite[Theorem 2]{rosen1965} to prove that the Nash equilibrium of $\mathcal{G}$, denoted by $\bs{\xi}$, is unique. 
%


\subsubsection{Electricity System Inefficiency}
\label{sec:inefficiency}

From the equilibrium conditions in \cref{eq:foc_optimal} and \cref{eq:foc_strategic} we know that 
\modification{
$\bs{\mu} \neq \bs{\xi}$. In other words, 
}
the user's individual interests are not aligned with the collective interests.
Particularly, the following result shows that users consume more resources in the NE.
\begin{theorem}\label{thm:efficiency}
 Suppose that Assumption \ref{as:1} is satisfied. Let $\bs{\mu}$ and $\bs{\xi}$ be the optimal and the Nash equilibrium (see \cref{eq:foc_optimal} and \cref{eq:foc_strategic}), respectively.
 Then, all users consume more resources in the NE, that is, $\xi_i \geq \mu_i$, for all $i \in \mathcal{P}$.
\end{theorem}

We use the concepts of the \emph{tragedy of the commons} and the \emph{price of anarchy} to analyze
the 
\modification{
\emph{efficiency loss}\footnote{Here we use the customer surplus as the efficiency metric; hence, the efficiency loss refers to the degradation of the customer surplus in the NE with respect to the optimal outcome.}
}
of the electricity system.
On one hand, the \emph{tragedy of the commons} is a situation in which the self interest of users leads to overuse of resources, despite of the harm to the society \cite{hardin1968tragedy}.
This occurs because 
the negative effects are shared by the community. 
Pitifully, the  benefit of the community decreases
when all individuals overuse resources.
We define the tragedy of the commons as follows.
\begin{definition}[Tragedy of the commons]
 At the optimal outcome, every agent has incentives to consume more resources.
\end{definition}
\modification{
Observe that the game $\mathcal{G}$ suffers the tragedy of the commons, since: i) every user has incentives to consume more than the optimal value of energy ($\xi_i\geq \mu_i$); and ii) the self interest leads to an undesired social outcome.
In particular, the next result shows that  users can consume twice the optimal demand in the NE
(when the price function $p(\cdot)$ is linear).
}
\begin{lemma}\label{res:convergence_xi}
Suppose that Assumption \ref{as:1} is satisfied and assume that 
 $\dot{C}(\cdot)$ is unbounded.
 If the users have the same valuation function with $\dot{v}_i(0) > p(0)$ and if the price function $p(\cdot)$ is linear, then the demand ratio $\slfrac{\normb{\xi}}{\normb{\mu}} \rightarrow 2 $ as $N\rightarrow \infty$.
 \end{lemma}

On the other hand, the
\emph{price of anarchy}
is the ratio between the worst possible NE and the social optimum \cite{papadimitriou2001algorithms, AlgorithmicG}.
 The next Theorem shows that the price of anarchy equals zero, that is, the customer surplus in the worst case NE is arbitrarily smaller than customer surplus in the social optimum.
 In contrast, systems with marginal cost prices, reduce their efficiency in at most $\slfrac{1}{4}$ \cite{johari2004efficiency}.

 %
 %
 %
 \begin{theorem}\label{res:boundary_efficiency}
 Suppose that Assumption \ref{as:1} is satisfied and assume that 
 $\dot{C}(\cdot)$ is unbounded
 . Let $\bs{\mu}$ and $\bs{\xi}$ be the optimal and the Nash equilibrium (see \cref{eq:foc_optimal} and \cref{eq:foc_strategic}), respectively.
 Then,
 the efficiency loss, that is,
the ratio of the customer surplus with the worst case NE and that of the optimal outcome, satisfies  
\begin{equation}\label{eq:ratio_aa}
r(\bs{\xi}, \bs{\mu}) = \frac{\sum_i v_i(\xi_i) - \normb{\xi} p(\normb{\xi}) }{ \sum_i v_i(\mu_i) - \normb{\mu} p(\normb{\mu}) }
\geq 0,
\end{equation}
with equality when all users have the same linear valuation with $\dot{v}_i(0) > p(0)$, $p(\cdot)$ is linear, and $N\rightarrow \infty$.
 \end{theorem}

\modification{ 
\begin{remark}
According to \cref{res:boundary_efficiency}, the worst case of efficiency loss  occurs when $N\rightarrow \infty$. Here we are not suggesting that the population grows to infinity, but that
the efficiency loss aggravates with the size of the population. 
\end{remark}

 

We can model the impact of inelastic loads by setting a minimum demand for each user (to account for the demand of appliances that do not react to prices). 
The total demand with these restrictions, denoted $||\hat {\boldsymbol{\mu}} ||$, may be higher than the  optimal demand without restrictions, that is,  $ || \boldsymbol{\mu}|| \leq || \hat {\boldsymbol{\mu}}|| \leq ||\boldsymbol{\xi}||$.
Thus, we can conjecture that  inelastic loads can reduce the efficiency loss, because they can reduce the distance between the optimal and the inefficient demand.
 }

\begin{example}
Let us illustrate loss of efficiency of the customer surplus as a function of the number of users $N$.
%
%
Here we use the following linear valuation function
\begin{equation}
 v_i (q_i) = \alpha_i q_i,
\end{equation}
where the parameter $\alpha_i$
 characterizes the valuation of the  $i\th$ agent (we select $\alpha_i=10$ for all $i\in\mathcal{P}$).
On the other hand, we define a quadratic generation cost function
$
 C(q) = \beta q^2 + b q,
$
and the unitary price function as
$
 p(g) = \sfrac{C(g)}{g} = \beta g + b,
$
where $\beta=1$ and $b=1$. Since all users have the same valuation function and  $\dot{v}_i(0) < p(0)$, then the individual demand is positive (see \cref{eq:foc_strategic}). 

\cref{fig:efficieny_loss} shows the ratio of demand and efficiency ratio as 
 a function of the number of users $N$.\footnote{\cref{sec:experiments} has details on the optimization method used in the experiments.}
On one hand, \cref{fig:demand_ratio} shows that the demand ratio $\normb{\xi}/\normb{\mu}$ closes to  $2$ for large populations, as expected from \cref{res:convergence_xi}.
On the other hand, \cref{fig:efficiency_ratio} shows that the efficiency ratio $r(\bs{\xi}, \bs{\mu})$ tends to zero, confirming the worst efficiency loss from \cref{res:boundary_efficiency}. 

%

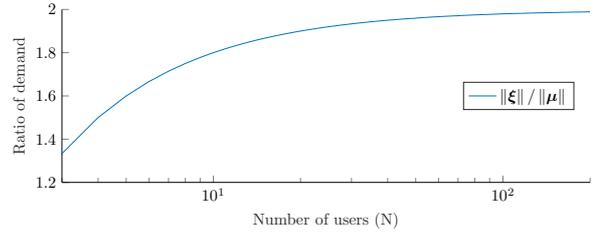
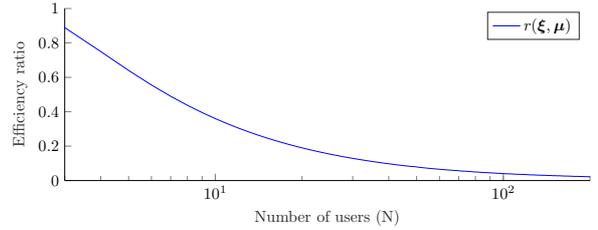
\begin{figure}
\centering
\begin{subfigure}[b]{\columnwidth}
 \resizebox{\columnwidth}{!}{
%
%
\definecolor{mycolor1}{rgb}{0.00000,0.44700,0.74100}%
\begin{tikzpicture}

\begin{axis}[%
width=4.521in,
height=1.493in,
at={(0.758in,2.554in)},
scale only axis,
xmode=log,
xmin=3,
xmax=200,
xminorticks=true,
xlabel style={font=\color{white!15!black}},
xlabel={Number of users (N)},
ymin=1.2,
ymax=2,
ylabel style={font=\color{white!15!black}},
ylabel={Ratio of demand},
axis background/.style={fill=white},
axis x line*=bottom,
axis y line*=left,
legend style={at={(0.97,0.5)}, anchor=east, legend cell align=left, align=left, draw=white!15!black}
]
\addplot [color=mycolor1]
  table[row sep=crcr]{%
2	1.00000007088372\\
3	1.33333340883495\\
4	1.49999994404843\\
5	1.60000032961566\\
6	1.66666657541744\\
7	1.71428575154037\\
8	1.75000094725681\\
9	1.77777805042702\\
10	1.80000208220501\\
11	1.81818176095734\\
12	1.83333379918859\\
13	1.8461537362419\\
14	1.85714277169038\\
15	1.8666673074557\\
16	1.87499947068297\\
17	1.88235403374787\\
18	1.88888887854599\\
19	1.89473619404259\\
20	1.89999969464342\\
21	1.90476254844517\\
22	1.90908922799921\\
23	1.91304291883626\\
24	1.91666609059727\\
25	1.92000099625257\\
26	1.92307776549661\\
27	1.92592615604217\\
28	1.92857209996283\\
29	1.93103497776663\\
30	1.93333242242313\\
31	1.93548339414253\\
32	1.9374820765886\\
33	1.93939200920295\\
34	1.94117434571577\\
35	1.94284897634023\\
36	1.94444359588783\\
37	1.94594511114637\\
38	1.94736690703988\\
39	1.9487084339652\\
40	1.94999776309281\\
41	1.95116708457842\\
42	1.95237934200402\\
43	1.95348834846113\\
44	1.95454556143224\\
45	1.95554525986926\\
46	1.95651543551474\\
47	1.95742269868608\\
48	1.95829502611958\\
49	1.95854246427121\\
50	1.95966182768698\\
51	1.96074180019044\\
52	1.96146043126715\\
53	1.96222577077115\\
54	1.96295631860942\\
55	1.9636324272933\\
56	1.96426789931952\\
57	1.96489569949636\\
58	1.96546780851875\\
59	1.96599949603014\\
60	1.96659469896558\\
61	1.96715394260071\\
62	1.96772314006279\\
63	1.96815918110048\\
64	1.96858619985368\\
65	1.96912724662409\\
66	1.96928610235889\\
67	1.96992561102787\\
68	1.97054161658913\\
69	1.970187806519\\
70	1.97121719411733\\
71	1.97156754960998\\
72	1.97214648232982\\
73	1.97237097989011\\
74	1.97257083768322\\
75	1.97281627157308\\
76	1.9733492695844\\
77	1.97361781211685\\
78	1.97404462720854\\
79	1.97449202380659\\
80	1.97486318226403\\
81	1.97486252108418\\
82	1.97520814014334\\
83	1.97566191310163\\
84	1.9759877823801\\
85	1.97606407465424\\
86	1.97640355380326\\
87	1.97678442845737\\
88	1.97652551512837\\
89	1.97699987022718\\
90	1.97749411882912\\
91	1.97728509808208\\
92	1.97793626600925\\
93	1.97778764253966\\
94	1.97830552486107\\
95	1.97859569342694\\
96	1.97874383844375\\
97	1.97855876823776\\
98	1.97927189775848\\
99	1.979396453019\\
100	1.97959180802395\\
101	1.97981422284317\\
102	1.97996722862801\\
103	1.98025834962102\\
104	1.98043833971101\\
105	1.9802124004913\\
106	1.98045294871734\\
107	1.98080812212432\\
108	1.98073113606942\\
109	1.98122221400571\\
110	1.98119250821415\\
111	1.98147644845965\\
112	1.98175725874448\\
113	1.98166379041446\\
114	1.98155167328834\\
115	1.98189182243533\\
116	1.98197334338185\\
117	1.98232055931783\\
118	1.98173706123342\\
119	1.98264890486511\\
120	1.98230666120947\\
121	1.9829115033643\\
122	1.98261413097977\\
123	1.98321658968394\\
124	1.98319527313475\\
125	1.98341575960938\\
126	1.98347251428262\\
127	1.98338064562949\\
128	1.98359316084519\\
129	1.98355153777514\\
130	1.98410461678861\\
131	1.98393309284839\\
132	1.98425419037221\\
133	1.98384344467177\\
134	1.98442621268864\\
135	1.98417974364191\\
136	1.98444416397119\\
137	1.98482951199933\\
138	1.98425445515854\\
139	1.98469469705369\\
140	1.98482155735983\\
141	1.98479973072445\\
142	1.98533987254837\\
143	1.98519312301563\\
144	1.98516134282223\\
145	1.98542666190311\\
146	1.98494211279896\\
147	1.98550231888859\\
148	1.98567656272286\\
149	1.98564121452724\\
150	1.98598941398158\\
151	1.9854486207643\\
152	1.98572704989093\\
153	1.9858514856876\\
154	1.98627602054639\\
155	1.98602823545041\\
156	1.98596406234843\\
157	1.98602747900899\\
158	1.98625098512787\\
159	1.98642638673758\\
160	1.9865407842733\\
161	1.98638921783175\\
162	1.98669816324354\\
163	1.98687895422297\\
164	1.98693120021928\\
165	1.98664269609981\\
166	1.98698493302859\\
167	1.98700542591061\\
168	1.98690033262677\\
169	1.9872097819263\\
170	1.98715575661612\\
171	1.98728576956771\\
172	1.98726441570441\\
173	1.98726656573499\\
174	1.98761794292177\\
175	1.98766959254541\\
176	1.98775573924014\\
177	1.98791786061063\\
178	1.98779157833998\\
179	1.98751225981567\\
180	1.98807593980891\\
181	1.9876332969551\\
182	1.98789287621517\\
183	1.98793493951298\\
184	1.98798806311122\\
185	1.98814832324239\\
186	1.9879257436828\\
187	1.98792260837502\\
188	1.98841495470993\\
189	1.98802225839123\\
190	1.98858080933422\\
191	1.98862595099478\\
192	1.98848797033974\\
193	1.98859873468245\\
194	1.98858545002164\\
195	1.98841116080848\\
196	1.98876405715219\\
197	1.98877187041122\\
198	1.98893129613913\\
199	1.98877541409639\\
200	1.9886735845313\\
};
\addlegendentry{$\normb{\xi}/\normb{\mu}$}

\end{axis}
\end{tikzpicture}%
 } 
 \caption{Ratio of demand $\normb{\xi}/\normb{\mu}$.}
 \label{fig:demand_ratio}
 \end{subfigure}
 
 \begin{subfigure}[b]{\columnwidth}
 \resizebox{\columnwidth}{!}{
%
%
\begin{tikzpicture}

\begin{axis}[%
width=4.521in,
height=1.493in,
at={(0.758in,0.481in)},
scale only axis,
xmode=log,
xmin=3,
xmax=200,
xminorticks=true,
xlabel style={font=\color{white!15!black}},
xlabel={Number of users (N)},
ymin=0,
ymax=1,
ylabel style={font=\color{white!15!black}},
ylabel={Efficiency ratio},
axis background/.style={fill=white},
axis x line*=bottom,
axis y line*=left,
legend style={legend cell align=left, align=left, draw=white!15!black}
]
\addplot [color=blue]
  table[row sep=crcr]{%
2	0.999999999999999\\
3	0.888888833214011\\
4	0.750000023047705\\
5	0.639999739072994\\
6	0.555555827371946\\
7	0.489795637395707\\
8	0.437498611913424\\
9	0.395061375447678\\
10	0.359997324296872\\
11	0.330578558704154\\
12	0.305554810121041\\
13	0.284023859874391\\
14	0.265306044908794\\
15	0.248889161329602\\
16	0.234375168234254\\
17	0.221451041777485\\
18	0.209876723937735\\
19	0.199446381111953\\
20	0.190000668843825\\
21	0.181405022007841\\
22	0.173556822652908\\
23	0.16635220671778\\
24	0.159723343525457\\
25	0.153598713997882\\
26	0.147927618047662\\
27	0.142660966378977\\
28	0.137752994610783\\
29	0.133173859492907\\
30	0.128890404327053\\
31	0.124870850109391\\
32	0.121127440054656\\
33	0.117543862520653\\
34	0.114190846863991\\
35	0.111035924806739\\
36	0.108026344222542\\
37	0.105188113827384\\
38	0.102495859165611\\
39	0.0999518684857658\\
40	0.0975039903563056\\
41	0.0952812656025547\\
42	0.0929733084115539\\
43	0.0908608339685357\\
44	0.0888454984664162\\
45	0.0869332502530537\\
46	0.0850779427686449\\
47	0.0833422873942934\\
48	0.0816719181236976\\
49	0.0811963111223872\\
50	0.0790500486575092\\
51	0.0769753653544321\\
52	0.0755992145264246\\
53	0.0741213434640625\\
54	0.0727139139653843\\
55	0.0714147595211412\\
56	0.0701882651280082\\
57	0.0689766501383387\\
58	0.0678715599521916\\
59	0.0668449632391421\\
60	0.0656951158206853\\
61	0.0646131811716464\\
62	0.0635137385387224\\
63	0.0626679816962302\\
64	0.0618422526871569\\
65	0.0607922752175305\\
66	0.0604828315700425\\
67	0.0592441883145934\\
68	0.0580494207134201\\
69	0.058730448582302\\
70	0.0567362995793021\\
71	0.0560565535813048\\
72	0.0549311364490668\\
73	0.054495411000049\\
74	0.054106365534186\\
75	0.053629006236216\\
76	0.0525914049612816\\
77	0.0520675461760104\\
78	0.0512372146894159\\
79	0.0503649565063223\\
80	0.0496405475208474\\
81	0.0496412483754439\\
82	0.0489683106947145\\
83	0.0480875735529589\\
84	0.0474477297582761\\
85	0.0472987291198081\\
86	0.0466353350134549\\
87	0.0458903352123168\\
88	0.0463979709565233\\
89	0.0454722209314877\\
90	0.0445055765340481\\
91	0.0449140294469037\\
92	0.0436403324802426\\
93	0.0439315074262825\\
94	0.0429174115210606\\
95	0.0423514098660804\\
96	0.0420604237874508\\
97	0.0424240002794597\\
98	0.0410268019982486\\
99	0.0407820337120274\\
100	0.0403971626589037\\
101	0.0399645906826912\\
102	0.0396643067497357\\
103	0.0390939715292545\\
104	0.0387417563916171\\
105	0.0391840176363518\\
106	0.038712705569488\\
107	0.0380182866742684\\
108	0.0381696017931704\\
109	0.0372026271201917\\
110	0.0372612564591892\\
111	0.0367039356131529\\
112	0.0361526357259479\\
113	0.0363375070345026\\
114	0.0365564457655908\\
115	0.0358879687087145\\
116	0.035728286003948\\
117	0.0350495697090139\\
118	0.0361924380680456\\
119	0.0344011299413492\\
120	0.0350735769459326\\
121	0.0338848639039666\\
122	0.0344706774816792\\
123	0.0332844202452869\\
124	0.0333268763352171\\
125	0.032893531248198\\
126	0.032781357277435\\
127	0.0329626648312344\\
128	0.0325445854688318\\
129	0.0326267845244433\\
130	0.0315376308779973\\
131	0.0318756110193092\\
132	0.0312442807515409\\
133	0.0320521399786556\\
134	0.0309080785651469\\
135	0.0313903330788834\\
136	0.0308697043761832\\
137	0.0301110016776983\\
138	0.0312432457888654\\
139	0.0303763670862131\\
140	0.0301263513725044\\
141	0.0301695465485868\\
142	0.0291040749310902\\
143	0.029394529364921\\
144	0.0294573189061568\\
145	0.0289331689398944\\
146	0.0298871558070619\\
147	0.0287854756369723\\
148	0.0284418208810673\\
149	0.0285134211338694\\
150	0.0278248688656413\\
151	0.0288906477219274\\
152	0.0283427845327327\\
153	0.028096872589162\\
154	0.0272592619142328\\
155	0.0277522912933178\\
156	0.0278757554561542\\
157	0.0277487285928477\\
158	0.0273094597673883\\
159	0.0269622977313091\\
160	0.0267367804942145\\
161	0.0270367881054862\\
162	0.026425717754057\\
163	0.0260704054944784\\
164	0.0259667579493549\\
165	0.0265355864037569\\
166	0.0258606453106688\\
167	0.0258203001308898\\
168	0.0260277704951519\\
169	0.0254152879105385\\
170	0.0255241277144442\\
171	0.0252662711853695\\
172	0.025309026611163\\
173	0.0253022581773013\\
174	0.0246112285972213\\
175	0.0245119901522319\\
176	0.0243395490635057\\
177	0.024015480598876\\
178	0.0242678568542837\\
179	0.0248205114859121\\
180	0.023707067116622\\
181	0.0245803252847996\\
182	0.0240676115553944\\
183	0.0239826956635314\\
184	0.0238789723604851\\
185	0.0235629078344193\\
186	0.0240154660256048\\
187	0.0240080550297553\\
188	0.0230353928586011\\
189	0.0238129737000814\\
190	0.0227079217543073\\
191	0.022618481326609\\
192	0.0228899948413543\\
193	0.0226720171796276\\
194	0.0226983482041386\\
195	0.0230521694855\\
196	0.0223457609659484\\
197	0.0223274083330517\\
198	0.0220132974697894\\
199	0.022323249065208\\
200	0.0225234121782372\\
};
\addlegendentry{$r(\bs{\xi}, \bs{\mu})$}

\end{axis}
\end{tikzpicture}%
 }
 \caption{Efficiency ratio $r(\bs{\xi}, \bs{\mu})$.}
 \label{fig:efficiency_ratio}
 \end{subfigure}

 \caption{Ratio of demand and efficiency ratio as a function of the number of users $N$. In the worst case the demand ratio tends to $2$ and the efficiency ratio (see \cref{eq:ratio_aa}) tends to $0$ 
 as $N\rightarrow \infty$.}
 \label{fig:efficieny_loss}
\end{figure}
\end{example}

 These results show that lack of coordination harms the efficiency of the system. In particular, the users can consume up to twice the optimal demand and obtain a small fraction of the optimal surplus.
 Next we propose an incentive scheme that mitigates the system's inefficiencies.

\section{Incentives-Based Mechanism}\label{sec:incentives}

The  efficiency loss of the Nash equilibrium motivates the design of an incentives mechanism to improve the equilibrium in strategic environments \cite{mas1995microeconomic, hurwicz06}.
%
In this section, we introduce a price mechanism that improves the efficiency of the system when users are strategic. Also, we show two instances of the mechanism with different properties.


From the perspective of mechanism design, 
the \emph{principal} (in this case the utility) can design
the payoff structure of the game to reach a desired social goal.
Hence, the principal intervenes to guarantee that selfish users reach an equilibrium that
maximizes the customer surplus; this considering the
private information held by each agent, 
necessary to calculate the optimal outcome.
%
%
Summarizing, mechanism design consists in designing a solution system to a decentralized optimization problem with private information
\cite{hurwicz06}. 
A mechanism $\mathcal{M}=\{ \Sigma_1, \ldots, \Sigma_N, o(\cdot) \}$ defines a set of strategies $\Sigma_i$ for each player and an outcome rule $ o:\Sigma_1 \times \ldots \times \Sigma_N \to \mathcal{O}$ that maps from the set of possible strategies  to the set  of possible outcomes $\mathcal{O}$ \cite{mas1995microeconomic}.
%

  \begin{figure}[tb]
  \centering
   \resizebox{\columnwidth}{!}{
 \begin{tikzpicture}[auto, thick, >=latex]
 
 \node[basic box b=black] (box1) at (0,0) {$g=\sum_i q_i$};
 
 \node [above] at (box1.north) {Utility};

 \path (box1) ++(120pt, 0) coordinate (b1);
 
 \path (b1) ++(60:3mm)  node[basic box b=black, text=white, fill=white] (box2a)  {$q_i \in \underset{x_i}{\argmax} U_i(x_i, g) + I(x_i, g) $};

 \path (b1) ++(60:1.5mm)  node[basic box b=black, text=white, fill=white] (box2b)  {$q_i \in \underset{x_i}{\argmax} U_i(x_i, g) + I(x_i, g) $}; 
 
 \node[basic box b=black, fill=white] (box2) at (b1)  {$q_i \in \underset{x_i}{\argmax} U_i(x_i, g) + I(x_i, g) $};

 \node [above] at (box2a.north) {Users};

 \draw [state, ->] (box1.east) to node[name=g] {$g$}  (box2.west);

 \def\h{1.2cm}
 
 \draw [rounded corners=7pt, <-] (box1.west) -- ++(-.5, 0) |- ++(0.5, -\h) node(p1) {};
  
 
 

 \path (box2.east) ++(0.5, -\h/2) ++(0:1.5mm) coordinate (i2);

 \draw [rounded corners=7pt, -] (box2b.east) -| (i2) |- ++(-.5, -\h/2)  node(p2) {};
 
 \draw[rounded corners=7pt, -] (p1.base) -- (p2.base) node[pos=.5] {$q_1, \ldots, q_N$}; 
 
 \end{tikzpicture}   
 }
   \caption{Scheme of a decentralized price mechanism. 
   The utility broadcasts the total demand so that each agent can determine its optimal demand.}
   \label{fig:decentralized}
 \end{figure}
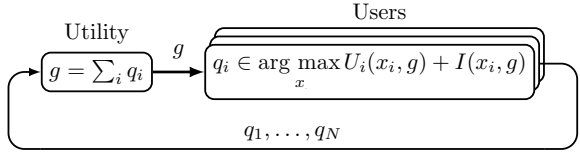

%
%
\modification{
%
Here we propose an \emph{indirect revelation mechanism}, which avoids the revelation of private information,
and
uses 
a one-dimensional message space 
(see Fig. \ref{fig:decentralized}).
}
In this mechanism, the users communicate their demand $q_i$ to the principal, who in turn calculates and broadcasts the total demand $\normb{q}$ to the users.\footnote{
We assume that the utility's infrastructure prevent users from tampering their electricity meters to misreport their demand. Nonetheless, 
the principal can calculate the total demand 
indirectly, inquiring the total generated energy. In other words, the principal can use the feedback of the physical world to gather information.}
Then, 
 users use the incoming information to
 adjust their demand.
In this mechanism the users have the same strategy space ($(\Sigma_i)_{i\in\mathcal{P}}$) as in the original system (game $\mathcal{G}$).

The outcome rule $o(\cdot)$ of the mechanism returns  the payment for each user;
hence, $o(\cdot) = [o_1(\cdot), \ldots, o_N(\cdot)]$ and 
the outcome space is $\mathcal{O} \equiv  \mathbb{R}_{\geq 0}^N$. 
The outcome rule $o(\cdot)$
must
guarantee that 
the game has a NE equal to the optimal outcome $\bs{\mu}$. 
To this end, we modify the payment function $t_i(\cdot)$ adding an incentive  $I_i(\cdot)$ that aligns the users' profit  with the population's objective function (customer surplus).
We achieve this by selecting a function $I_i(\cdot)$ that estimates the externalities (or the impact on prices) caused by the $i^{th}$ user.

Let the payment of the $i\th$ user be
\begin{equation}\label{eq:o_i}
o_i(\bs{q}) = t_i(\bs{q}) + I_i(\bs{q}),
\end{equation}
 with
 incentives of  the form
\begin{equation}\label{eq:I_i}
I_i(\bs{\bs{q}}) = I(q_i, \normb{q}) = \norm{\bs{q}_{-i}}  \left( h(\norm{\bs{q}_{-i}})  - p\left( \normb{q} \right) \right),
\end{equation}
where 
$\norm{\bs{q}_{-i}} = \normb{q} - q_i$ and the function
$h(\norm{\bs{q}_{-i}})$ estimates the electricity price without the $i\th$ user in the  system. 
We choose this incentive structure because each user only knows its own demand $q_i$ and the total demand $\normb{q}$, ignoring the precise demand of other users. 
Furthermore, we assume that 
users are indistinguishable in the system; hence, 
two users with the same demand impose the same externalities to the system.
%
%
The form of this incentive is related to the price used in the VCG
mechanism \cite{AlgorithmicG}
and some  payoff functions used in potential games \cite{monderer1996potential}.  
%

Introducing the incentives in  \cref{eq:I_i} we obtain a new game 
defined as the 3-tuple  $\mathcal{G}_\mathcal{I} = \langle \mathcal{P}, (\Sigma_i)_{i \in \mathcal{P}}, (W_i)_{i \in \mathcal{P}} \rangle$, where $\mathcal{P}$ is the set of players, $\Sigma_i$ is the set of available strategies of each player, and $W_i: \Sigma_1 \times \dots \times \Sigma_N \to \mathbb{R}$ is the surplus function of the $i\th$ player,  defined as
\modification{
\begin{multline}
W_i(q_i,\bs{q}_{-i}) = v_i(q_i) + o_i(\bs{q}) =  U_i( q_i,\bs{q}_{-i} ) + I_i( \bs{q} )
\\ = v_i(q_i) -  \normb{q} p\left( \normb{q} \right) + || \bs{q}_{-i} || h_i(\norm{\bs{q}_{-i} }).
\label{eq:W_i}
\end{multline}
}
Observe that the NE of the game $\mathcal{G}_\mathcal{I}$ 
has the same equilibrium conditions than 
\modification{
the optimal outcome in \cref{sec:optimal_equilibrium};
}
hence, the NE of $\mathcal{G}_\mathcal{I}$ is equal to the optimal equilibrium $\bs{\mu}$.

\subsection{Mechanism's Properties}

Now we analyze the properties of the incentive mechanism and provide of two instances that have different properties. First, let us define some desirable properties of mechanisms, namely \emph{budget balance} and \emph{individual rationality}.

\subsubsection{Budget Balance}

It is desirable to make   
the net payments equal to zero, that is, 
guarantee that the sum of charges is equal to the total cost \cite{AlgorithmicG}. 
In this way the system doesn't require external subsidies nor imposes taxes on the population.
This property is known as budget balance, and is defined as follows
\begin{definition}
 A mechanism that implements payments $o_i(\cdot)$ is budget balanced 
 if
$
 \sum\nolimits_{i\in\mathcal{P}} o_i(\bs{q}) = C(\norm{\bs{q}}).
$
\end{definition}

Note that the original game ($\mathcal{G}$) satisfies the budget balance property, because the total payments equal the generation cost (see  \cref{eq:avg_cost}).
Therefore, we need that
$ \sum\nolimits_{i\in\mathcal{P}} I_i(\bs{q}) = 0$
to preserve the budget balance in $\mathcal{G}_\mathcal{I}$.
%
%
%
%
However, the next theorem shows that it is impossible to find a function $h(\cdot)$ that 
balances the amount of rewards (price discounts) and penalties (price increment) imposed to customers. Consequently,  the mechanism requires either inflow or outflow of resources.
%
\begin{theorem}\label{thm:budget}
Suppose that Assumption \ref{as:1} is satisfied and let $p(\cdot)$ be the average cost price function. Then
there is no function $h(\cdot)$ such that a mechanism with payments defined by \cref{eq:o_i}  satisfies the budget balance property.
\end{theorem}

This result is an analogous to the Myerson-Satterthwaite impossibility theorem \cite{Myerson83}, which states
the impossibility of designing a mechanisms with ex-post efficiency and
without external subsidies in games between two parties. However, \cref{thm:budget} considers a particular case with 1) nonlinear prices (subject to \cref{as:1}) and 2) the customer surplus as the  maximization criteria, rather than the aggregate surplus.


\subsubsection{Individual Rationality}

The utility company cannot force users to sign contracts accepting the incentives scheme. Hence, the mechanism must ensure that every user voluntarily decides to join the system with incentives (game $\mathcal{G}_\mathcal{I}$). In particular, a rational user would accept a mechanism if it guarantees higher benefits participating.
\modification{
This property
}
 is called \emph{individual rationality} and is defined as
\begin{definition}
A mechanism is individual rational if the surplus at the equilibrium is larger than zero, that is,  $W_i(\bs{\mu})\geq 0$.
\end{definition}
The following result shows the conditions to guarantee that the proposed mechanism satisfies the individual rationality.
\begin{proposition}\label{res:individual_rationality_cond}
Let \cref{as:1} be satisfied.
Consider a mechanisms with payments given by \cref{eq:o_i}. 
If $ h( \norm{ \bs{\mu}_{-i} } ) - p( \norm{ \bs{\mu}_{-i} }) \geq 0$, then the mechanism is individually rational.
\end{proposition}

\subsection{Examples of Incentives}
Here we present two instances of individual rational incentives that satisfy either budget deficit or surplus.
On one hand, 
the following result shows an instance of a mechanism that has budget deficit, that is, $\sum_{i\in \mathcal{P}} o_i(\bs{q}) \leq 0$. This implies that the mechanism requires external subsidies to operate.
\begin{proposition}\label{res:h_b}
Let \cref{as:1} be satisfied. A mechanism with payments given by \cref{eq:o_i} with 
\begin{equation}\label{eq:h_b}
h(g) = p\left(  g \right)
\end{equation}
satisfies the individual rationality property and has \emph{budget deficit}, i.e.,   $\sum_{i\in \mathcal{P}} o_i(\bs{q})
\leq C(\norm{\bs{q}})$ (or equivalently $\sum\nolimits_{i\in\mathcal{P}} I_i(\bs{q}) \leq 0$).
\end{proposition}

On the other hand, the following result shows a mechanism that  has budget surplus, that is, $\sum_{i\in \mathcal{P}} I_i(\bs{q}) \geq 0$, which implies that the mechanism imposes taxes on users.
\begin{proposition}\label{res:h_a}
Let \cref{as:1} be satisfied. A mechanism with payments  given by \cref{eq:o_i} with 
\begin{equation}\label{eq:h_a}
h(g) = p\left( \frac{N}{N-1} g \right)
\end{equation}
satisfies the individual rationality property and has \emph{budget surplus}, i.e.,  $\sum_{i\in \mathcal{P}} o_i(\bs{q}) 
\geq C(\norm{\bs{q}})$ (or equivalently $\sum\nolimits_{i\in\mathcal{P}} I_i(\bs{q}) \geq 0$).
\end{proposition}



\cref{tab:summary_properties} shows a summary of the properties of incentives with $h(\cdot)$ given by \cref{eq:h_b} and \cref{eq:h_a}, denoted $I^d(\cdot)$ and $I^s(\cdot)$, respectively.
%
%
\begin{table*}
\centering
\caption{Properties of the mechanisms.}
\begin{tabular}{@{}p{.25\textwidth}cp{.28\textwidth}cp{.28\textwidth}@{}} \toprule
		&  &   $h(g) = p(\frac{N}{N-1} g)$   &  &   $h(g) = p(g)$ \\ \midrule
If $q_i=q_j$, $\forall \, i,j\in\mathcal{P}$   &  &   $I^s_i(\bs{q}) = 0$  &  & $I^d_i(\bs{q}) \leq 0$\\   
Weak budget balance   & &  $\sum_{i\in\mathcal{P}} I^s_i(\bs{q}) \geq 0$  & &  $\sum_{i\in\mathcal{P}} I^d_i(\bs{q}) \leq 0$ \\
Individual rationality & & $W_i(\bs{\mu}) \geq 0$ & & $W_i(\bs{\mu}) \geq 0$ \\
\bottomrule
\end{tabular}
\label{tab:summary_properties}
\end{table*}
%

\begin{example}
Let us illustrate 
the effect of incentives in the customer surplus as a function of the number of users.
In this case we use nonlinear valuation functions of the form
\begin{equation}
 v_i (q_i) = \alpha_i \log(1+q_i),
\end{equation}
with 
\begin{equation}
\alpha_i = 10+\frac{i-1}{N-1}.
\end{equation}

\modification{
If the unitary price function is
$
 p(g) = \sfrac{C(g)}{g} = \beta g + b,
$
then the incentives from \cref{tab:summary_properties} are
\begin{equation}
 I_i^s(\bs{q}) = \norm{ \bs{ q }_{-i} } \frac{\beta}{N-1} \left( \norm{ \bs{q}_{-i} } - q_i(N-1) \right)
\end{equation}
and 
\begin{equation}
 I_i^d(\bs{q}) = - \norm{ \bs{ q }_{-i} } \beta q_i.
\end{equation}
}

\cref{fig:surplus} shows that the 
 mechanism with budget deficit ($I^d$) generates less profit for the society, but the profit is positive (because the mechanism is individual rational). Furthermore, the mechanism with budget surplus ($I^s$) does not have a notorious effect in the customer surplus (in the simulations $0 \leq \sum_{i\in\mathcal{P}}  I_i^s(\bs{q}) \leq 0.0983$).
%
%
\begin{figure}
\centering
 \resizebox{\columnwidth}{!}{
 \input{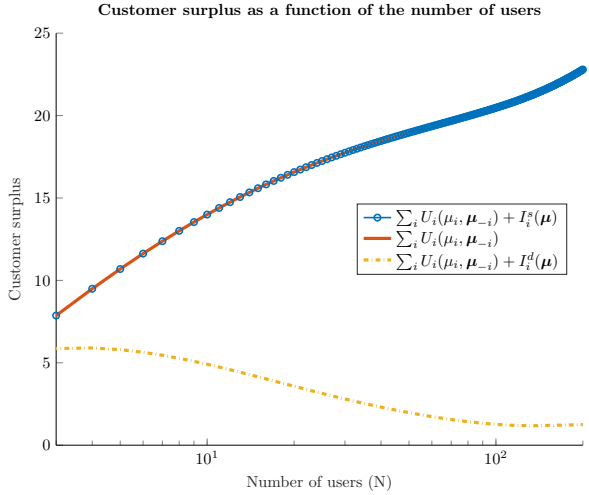}  
 }
 \caption{Customer surplus with and without incentives for different population sizes. The mechanism with budget deficit generates less profit for the users, but the profit is positive.}
 \label{fig:surplus}
\end{figure}

\end{example}

\section{Conclusions and Future Directions}
\label{sec:concl}

\modification{
In this work we find that, without a proper coordination, users who become strategic (thanks to new technologies) can harm the efficiency of the power system. 
}
Particularly, with the average cost prices, the efficiency loss can be arbitrarily large, while systems that use marginal cost prices have an efficiency loss of at most $\slfrac{1}{4}$ \cite{johari2004efficiency}.
Furthermore, in the worst case NE the users consume twice the optimal energy.

We propose an indirect revelation mechanism to improve the efficiency of the system.
\modification{
Our mechanism uses a one dimensional message space and avoids revelation of private information.}
We show two instances of the mechanism that satisfy either budget deficit or budget surplus. Furthermore, the properties of the mechanism guarantee that  users would join the incentives program voluntarily, since they have non-negative surplus.

In this work, we use a stylized power system model to facilitate the analysis,
however, 
it is important to consider
 the stochastic nature of both demand and generation and
temporal interdependencies of the demand.

\section*{Acknowledgements}

The paper has benefited substantially from conversations with Ramesh Johari and the comments from anonymous reviewers.
This work has been supported in part by Project SGR C/marca 2015.

\section{Appendix}

In this section we include the proofs of the results introduced in the paper.

\subsection{Optimal Price Function in Non-strategic Settings}

The following proof demonstrates that the average cost price function maximizes the customer surplus when users are non-strategic.

\checkl{thm:avg_prices}{
\begin{proof}[Proof of \cref{thm:avg_prices}]
The proof has two steps. First, we show that 
%
the system's equilibrium $(\tilde p, \tq)$, with 
$\tilde{p} = C\left( \norm{\tq} \right)/ \norm{ \tq }$
(the average cost evaluated at the efficient outcome $\tilde{\bs{q}}$)
maximizes the customer surplus of non-strategic users (see \cref{eq:opt_price_takers}).
%
%
However,  the optimal price $\tilde p$, which is constant, depends on the optimal allocation, which is unknown beforehand. 
In the second step we show that a power system with the average cost price function 
$
p(\normb{q}) = \slfrac{ C\left( \normb{q} \right) }{ \normb{ q } }
$
clears with the efficient allocation $\tq$. 
In other words, 
with the average cost price function the system's equilibrium maximizes the customer surplus.

\textbf{Step (i):}  
Let us define 
the restrictions of the optimization problem in \cref{eq:opt_price_takers} as 
\begin{equation}\label{eq:res_cost}
\Psi(\bs{q}, p) = C\left( \normb{q} \right) - \normb{q} p
\end{equation}
and
\begin{equation}\label{eq:res_users}
\Phi_i(q_i, p) = q_i(\dot v_i(q_i) - p),
\end{equation}
where $(p, \bs{q})$ is a system's equilibrium.
\modification{
The Lagrangian of the problem in \cref{eq:opt_price_takers} is 
}
\begin{multline}\label{eq:laplacian_a}
 \mathcal{L}(\bs{q}, p, \bs{\lambda}, \bs{\eta}) = \sum\nolimits_{i} \tilde{U}_i (q_i, p) - 
 \lambda_0 \Psi(\bs{q}, p)  \\
 - \sum\nolimits_i \eta_i \Phi_i(q_i, p) + \sum\nolimits_i \lambda_i q_i,
\end{multline}
where $\bs{\lambda}=[\lambda_0, \ldots, \lambda_N]$ and $\bs{\eta} = [\eta_1, \ldots, \eta_N]$.
The tuple that maximizes the customer surplus, denoted $(\tilde{p}, \tq)$, satisfies the following 
first order necessary conditions
\cite{kuhn1951proceedings, karush1939minima}
\begin{equation}\label{eq:cond_p}
 \pde{ \mathcal{L}(\tq, \tilde{p}, \bs{\lambda}, \bs{\eta})  }{p} = \sum\nolimits_i \tilde{q}_i (\lambda_0 - 1) + \sum\nolimits_i \eta_i \tilde{q}_i  = 0,
\end{equation}
\begin{equation}\label{eq:cond_qi}
 \pde{ \mathcal{L}(\tq, \tilde{p}, \bs{\lambda}, \bs{\eta})  }{q_i} = \phi_i
 - \lambda_0 \psi 
- \eta_i ( \phi_i + \tilde q_i \ddot{v}_i(\tilde{q}_i) ) + \lambda_i  
  = 0,
\end{equation}
\begin{equation}\label{eq:eq_restriction}
\Phi_i(\tilde q_i, \tilde{p})  = 0,
\end{equation}
\begin{equation}\label{eq:feasible_cond}
  \lambda_0 \Psi(\tq, \tilde{p}) = 0,
\end{equation}
\begin{equation}\label{eq:greater_zero_cond}
\lambda_i \tilde{q}_i = 0,
\end{equation}
with $\phi_i = \dot{v}_i (\tilde{q}_i) - \tilde{p}$, $\psi = \dot{C}\left( \norm{\tq} \right)  - \tilde{p}$,  $\lambda_k\geq0$,  and $\eta_i\geq 0$, for 
$i\in\{1, \ldots, N\}$ and $k\in\{0, \ldots, N\}$.

Let us assume by contradiction that the cost restriction \cref{eq:res_cost} does not bind. Therefore, $\Psi(\tq, \tilde p) \leq 0$, 
\modification{
which implies that $ \tilde p \geq C\left( \norm{\tq} \right) / \norm{\tq} > 0$ (the rightmost inequality follows because $C(\norm{\tq}) > 0$ if $\norm{\tq} > 0$).
}
Thus,  we need $\lambda_0=0$ to satisfy \cref{eq:feasible_cond}. Moreover, for every user $i$ with $\tilde q_i > 0$ we have $\lambda_i = 0$ and $\phi_i=0$ to satisfy \cref{eq:greater_zero_cond} and $\Phi_i(\tilde q_i, \tilde p) = 0$ (see the equality constraint in \cref{eq:res_users}).
Also, when $\tilde q_i > 0$, $\eta_i$ must be positive to satisfy \cref{eq:cond_p}.

With the previous considerations \cref{eq:cond_qi} becomes
\begin{equation}\label{eq:contradicting}
 \eta_i \tilde q_i \ddot v_i (\tilde q_i) = 0.
\end{equation}
\modification{
Observe that if the valuation function $v_i(\cdot)$ is concave, then $\ddot v_i (\tilde q_i) \leq 0$. Thus, \cref{eq:contradicting} holds only if $\ddot v_i (\tilde q_i) = 0$; however, since 
 $\phi_i=0$, we know that  $\dot{v}_i (\tilde{q}_i) = \tilde{p} = 0 $. 
 This leads to a contradiction, because  
 $ \tilde p \geq C\left( \norm{\tq} \right) / \norm{\tq} > 0$.
}
 In conclusion, the cost constraint in \cref{eq:res_cost} must bind in the equilibrium that maximizes the customer surplus. Hence, the price at the equilibrium is equal to the average cost.

\textbf{Step (ii):} Now, let us use the average cost pricing function $p(\normb{q}) = C\left(\normb{q}\right) / \normb{q}$ instead of a constant price. Here we intend to prove that the system has only one equilibrium, which corresponds to the Pareto optimal allocation.

Let us assume by contradiction that the system has two equilibria, namely the optimal allocation $\tq$ and a different allocation $\hat{\bs{q}}$
with 
$\tilde{q}_i \neq \hat{q}_i$ for some $i\in \mathcal{P}$.
First, let us assume that $\tilde{q}_i>\hat{q}_i>0$, then, 
$
 \dot{v}_i (\tilde{q}_i) < \dot{v}_i (\hat{q}_i) .
$
Our previous hypothesis, together with the optimal response of users 
(see \cref{eq:foc_price_takers}) and the average cost pricing, 
leads to
\begin{equation}\label{eq:contradicting_eq}
 \dot v_i (\tilde q_i) = \tilde p (\norm{\tq}) < \dot v_i (\hat q_i) = \hat p ( \norm{\hq} ).
\end{equation}
Since all users have the same marginal valuation, the previous expression holds for every user with positive demand. This implies that $\tilde q_i > \hat q_i$ for every
$i$ with $\hat q_i > 0$;
hence, we conclude that $\norm{\tq} > \norm{\hq}$.
Observe that \cref{eq:contradicting_eq} is true if the tariff $p(\cdot)$ strictly decreases with the total consumption, which contradicts \cref{as:1}. 

Hence, our initial assumption that $\tilde{q}_i > \hat{q}_i$ is false. Furthermore, we can construct a similar argument to show that $\tilde{q}_i < \hat{q}_i$ is false as well. Therefore, the system with the average pricing function allows only one equilibrium, which corresponds to the efficient allocation.
\end{proof}
}

\subsection{Efficiency Loss in Strategic Settings}

This proof shows that the optimal equilibrium and the NE in strategic settings are distinct; hence, this result shows that the system's NE is inefficient.
The next section investigates the efficiency loss in the NE with respect to the optimal equilibrium.

\checkl{thm:efficiency}{
\begin{proof}[Proof of \cref{thm:efficiency}]
\modification{
In this proof we show first that $||\boldsymbol{\mu}|| \leq ||\boldsymbol{\xi}||$ and then we prove that 
$\mu_j \leq \xi_j$ for all $j$.
}

First, let us assume by contradiction that 
 $\normb{\xi} < \normb{\mu}$, hence, there exist some $j$ such that $\mu_j > \xi_j>0$ with the following equilibrium condition (see \cref{eq:foc_optimal}) 
 \begin{equation}\label{eq:condition_contradiction}
 \dot{v}_j(\mu_j) = 
\dot{C}(\normb{\mu})  \geq  \dot{C}(\normb{\xi}), 
 \end{equation}
 where $\dot{C}(g) = p(g) + g\dot{p}(g)$ (see \cref{eq:avg_cost}).
We can use the equilibrium conditions in \cref{eq:foc_strategic} to express the previous inequality as
\begin{equation}\label{eq:condition_contradiction_b}
\dot{v}_j(\mu_j) \geq \dot{v}_j(\xi_j) + (\normb{\xi} - \xi_j) \dot{p}(\normb{\xi}).
\end{equation}
 Since $\mu_j>\xi_j$, then $\dot{v}_j(\mu_j) < \dot{v}_j(\xi_j)$, because $v_i(\cdot)$ is concave and increasing; hence, \cref{eq:condition_contradiction_b} is true if $(\normb{\xi} - \xi_j) \dot{p}(\normb{\xi}) < 0$, which is not possible, since $\normb{\xi} - \xi_i \geq 0$ and $\dot{p}(\cdot) > 0$ (see \cref{as:1}). 
 From this contradiction we conclude that 
 $\normb{\xi} \geq \normb{\mu}$.
 
 Now, let us assume by contradiction that there exists some $j$ such that $\mu_j> \xi_j$. 
 With the same argument used at the beginning of the proof we conclude that
 \begin{multline}
 \dot{v}_j(\mu_j) = \dot{C}( \normb{\mu} )  \leq 
\dot{C}( \normb{\xi} ) 
 \\ = 
\dot{v}_j(\xi_j) +  (\normb{\xi} -\xi_j) \dot{p}(\normb{\xi}).
 \end{multline}
 However, the previous expression requires that $(\normb{\xi} -\xi_j) \dot{p}(\normb{\xi}) < 0$, which is not possible. Therefore, we conclude that $\mu_j\leq \xi_j$ for all $j\in\mathcal{P}$. 
\end{proof}
}

\subsection{Price of Anarchy}

Here we undertake the task of investigating the price of anarchy when users become strategic. 
First,  we analyze how the total demand changes as the number of users increases. Later, we find the conditions in which the ratio of demand $\sfrac{ \normb{\xi} }{ \normb{ \mu } }$ reaches the upper bound. These results allow us to prove that the efficiency loss is arbitrarily large.

The first result shows that the total demand of the system cannot decrease  as the number 
of users increases.
In particular, it can occur that only a subset of users with high valuations can afford the electricity prices (see \cref{eq:foc_strategic}).
Hence, the demand does not necessarily increase as the population grows. 
\begin{proposition}\label{prop:inc_demand}
Suppose that Assumption \ref{as:1} is satisfied and let $\bs{\xi}$ and $\hat{\bs{\xi}}$ be the Nash equilibrium of two systems with  $\mathcal{P}$ and $\hat{\mathcal{P}}$ users, respectively. If $\mathcal{P} \subseteq \hat{\mathcal{P}}$, then $\normb{\xi} \leq \norms{\hat{\bs{\xi}}} $.
%
\end{proposition}

\checkl{prop:inc_demand}{
\begin{proof}
We assume by contradiction that the total demand in the NE strictly decreases with the number of users, that is,  $\normb{\xi} > \norms{\hat{\bs{\xi}}} $. 
Hence, 
at least one user $j\in\mathcal{P}$ reduces its demand, that is,  there exists some $j$ such that $\xi_j>\hat{\xi}_j > 0$.
From \cref{eq:foc_strategic} we know that $\hat{\xi}_j$ satisfies
\begin{equation}\label{eq:bound_v_xi_n_1}
\dot v_j (\hat \xi_j) = \dot{t}_j(\hat{\bs{\xi}}) =  p\left( \norms{\hat{\bs{\xi}}} \right) + \hat \xi_j \dot p \left( \norms{\hat{\bs{\xi}}} \right).
\end{equation}
The right hand side of \cref{eq:bound_v_xi_n_1} is increasing with respect to $\hat \xi_j$ (see \cref{as:1}).
Therefore, 
we can find an upper bound of $\dot{v}_j(\hat \xi_j)$ replacing in the right hand side of \cref{eq:bound_v_xi_n_1} $\hat \xi_j$ by $k_j = \delta + \hat \xi_j$, where $\delta =  \norm{{\bs{\xi}}} - \norms{\hat{\bs{\xi}}} > 0$ (the inequality follows from our initial assumption), resulting
%
\modification{
\begin{equation}\label{eq:bound_v_xi_n_2}
\dot v_j (\hat \xi_j) < p\left( \norm{{\bs{\xi}}} \right) + \hat \xi_j   \dot p \left( \norm{{\bs{\xi}}} \right) + \delta \dot p \left( \norm{{\bs{\xi}}} \right). 
\end{equation}
Now, we can use our hypothesis  $\xi_j > \hat \xi_j$ and \cref{eq:bound_v_xi_n_1} to obtain an upper bound of \cref{eq:bound_v_xi_n_2}
\begin{equation}\label{eq:bound_v_xi_n_3}
p\left( \norm{{\bs{\xi}}} \right) + \hat \xi_j   \dot p \left( \norm{{\bs{\xi}}} \right) < p\left( \norm{{\bs{\xi}}} \right) + \xi_j   \dot p \left( \norm{{\bs{\xi}}} \right) = \dot v_j(\xi_j).
\end{equation}
Thus, from \cref{eq:bound_v_xi_n_3,eq:bound_v_xi_n_2} we have
\begin{equation}
\dot v_j( \hat \xi_j ) < \dot v_j (\xi_j) + \delta \dot{p} \left( \norm{{\bs{\xi}}} \right).
\end{equation}
}

From \cref{as:1} we know that 
 $\dot{p}(\normb{\xi}) > 0$. Since  $\delta >0$, \cref{eq:bound_v_xi_n_2}
implies that $\hat \xi_j > \xi_j$, contradicting our initial assumption.
%
%
%
Therefore we conclude that  the total demand cannot decrease with the number of users, i.e., $\normb{\xi} \leq \norms{\hat{\bs{\xi}}} $. 
%
%
%
\end{proof}
}

The following result shows some conditions to guarantee that all users have a positive demand.
\begin{lemma}\label{lemma:positive_demand}
 Suppose that Assumption \ref{as:1} is satisfied.
If all users have the same valuation function ($v_i(x)=v(x)$ for all $x\geq 0$)
 and $\dot{v} (0) > p \left( 0 \right)$, then they will have positive demand in the equilibrium, i.e., $\xi_i > 0$ for all $i\in\mathcal{P}$.
\end{lemma}

\checkl{lemma:positive_demand}{
\begin{proof}
If all users have the same valuation function, then they will have the same demand, that is, 
$\xi_i=\xi$ for all $i\in\mathcal{P}$ and $|| \bs{\xi} || = N\xi$. 
Therefore, the equilibrium conditions in \cref{eq:foc_strategic} become
\begin{equation}\label{eq:foc_strategic_same_val}
\begin{cases}
\xi = 0 
  & \quad \text{if }  \dot{v} (0) <  p \left( 0 \right), \\
 \dot{v} (\xi) = p \left( N \xi \right) + \xi \dot{p}\left( N\xi \right) & \quad \text{Otherwise}.
\end{cases} 
\end{equation}
Observe that  $\dot{v}(\xi)$ is decreasing and $p(N\xi) + \xi \dot{p}(N\xi)$ is increasing (see \cref{as:2}). Therefore, 
if $\dot{v} (0) > p \left( 0 \right)$, then there exists some $\xi>0$ such that 
$\dot{v} (\xi) = p \left( N \xi \right) + \xi \dot{p}\left( N\xi \right)$.
\end{proof}
}

In order to analyze the case in which every user has a positive demand, we assume that all users have the same valuation with $\dot{v}_i(0) > p(0)$ for all in $i\in \mathcal{P}$.
In this way we guarantee that $\xi_i>0$ (see \cref{lemma:positive_demand}).
The next Corollary shows that 
under the previous conditions 
the total demand strictly increases with the size of the population $N$.
\begin{corollary} \label{cor:strict_ineq}
Consider the conditions of \cref{prop:inc_demand}. 
If all users have identical valuation functions with $\dot{v}_i(0) > p(0)$, 
then the total  demand in the NE strictly increases with the number of users, that is, 
$\normb{\xi} < \norms{\hat{\bs{\xi}}}$. 
\end{corollary}

\checkl{cor:strict_ineq}{
\begin{proof}
Let  $\mathcal{P}=\{1, \ldots, N\}$ and $\hat{\mathcal{P}} = \mathcal{P} \cup \{N+1\}$.
Let all the users have the same valuation function, that is, 
$v_i(x) = v_j(x) = v(x)$ for all $x\in \mathbb{R}$ and $i, j \in \hat{\mathcal{P}}$. Consequently, all users will have the same demand, in other words, $\xi_i = \xi$ $i \in \mathcal{P}$  and $\hat \xi_j = \hat \xi $ for all $j \in \hat{\mathcal{P}}$ 
%
%
and the total demand for each system will be equal to
$\normb{\xi} = N \xi$ and $\norms{ \hat{ \bf{\xi} } } = (N+1)\hat \xi$.
 Furthermore, if $\dot{v}(0)>p(0)$, then $\xi, \hat \xi \neq 0$ (see \cref{lemma:positive_demand}).

Let us assume by contradiction that both systems have the same demand, i.e.,  $\normb{\xi} = \norms{ \hat{ \bs{\xi} } } = g$, which leads to $\frac{N}{N+1} \xi = \hat \xi$. 
Therefore, 
the larger population consumes les energy, 
that is, 
$
 \xi > \hat \xi.
$
%
 %
Hence, from the equilibrium conditions in \cref{eq:foc_strategic} we get
\begin{equation}
\dot v(\hat \xi) = p(g) + \hat{\xi} \dot p(g ) < 
p(g) + \xi \dot p(g ) = \dot v(\xi). 
\end{equation}
The previous expression implies that 
$\hat \xi$ is greater than $\xi$, which contradicts our initial hypothesis. 
Therefore, $\normb{\xi} \neq \norms{ \hat{ \bs{\xi} } }$, and from \cref{prop:inc_demand}
we conclude that 
when all users have the same valuation 
the total demand strictly increases with the number of users, that is, 
$\normb{\xi} < \norms{ \hat{ \bs{\xi} } }$.
\end{proof}
}

Now, we show that 
if the marginal cost function $\dot{C}(\cdot)$ is unbounded
then the total demand reaches a boundary as the population grows.
If the cost function is bounded, then  users would consume any amount of energy without paying significant prices.
\begin{lemma}\label{lemma:reducing_demand}
Suppose that Assumption \ref{as:1} is satisfied.
If $\dot{C}(\cdot)$ is unbounded, then the total demand reaches an upper bound as the number of users increase. That is,  $\normb{\mu} \rightarrow K_o$ and $\normb{\xi} \rightarrow K_s$ as $N\rightarrow \infty$, for $K_o, K_s\in\mathbb{R}$.
Moreover,  if all users have the same valuation function with $\dot{v}_i(0) > p(0)$, then the individual demand decreases as the population grows, that is,  $\mu_i\rightarrow 0$ and
$\xi_i\rightarrow 0$ as $N\rightarrow \infty$.
\end{lemma}

\checkl{lemma:reducing_demand}{
\begin{proof}
%
%
Let us assume by contradiction that the optimal aggregate demand $\normb{\mu}$ is unbounded, that is, $\normb{\mu} \rightarrow \infty$ as $N\rightarrow \infty$. 
Hence, for every user $i$ with positive demand $\mu_i>0$  
the following equilibrium condition follows from \cref{eq:foc_optimal} 
\begin{equation}
\lim_{N\rightarrow\infty} \dot{v}_i(\mu_i) = \lim_{N\rightarrow\infty} \dot{C}(\normb{\mu}).
\end{equation}
If $\dot{C}(\cdot)$ is unbounded, then 
\begin{equation}\label{eq:unbounded}
\lim_{g\rightarrow\infty} \dot{C}(g) \rightarrow \infty.
\end{equation}
Hence, our initial hypothesis imply that $\dot{v}_i(\mu_i)\rightarrow \infty$ as $N\rightarrow \infty$,
 which contradicts \cref{as:1} because $v(\cdot)$ is not differentiable. 
 Therefore, we conclude that 
if $\dot{C}(\cdot)$ is unbounded, then the total demand  is bounded, that is, there exists some $K_o\in\mathbb{R}$ such that 
$\normb{\mu}\rightarrow K_o$ as $N\rightarrow \infty$.
Moreover, we can use a similar argument to prove that the aggregate demand in the NE $\left(\normb{\xi}\right)$ is also upper bounded by some positive number $K_s$.


Besides, 
if all users have the same valuation function, i.e., $v_i(x) = v_j(x) = v(x)$ for all $x\in \mathbb{R}$ and $i, j \in {\mathcal{P}}$, then they have the same demand, that is,  $\xi_i = \bar{\xi}$ and  $\mu_i = \bar{\mu}$ for all $i\in\mathcal{P}$. 
%
If 
the total demand is bounded, then 
\begin{equation}
\lim_{N\rightarrow \infty} \bar{\mu} =  \lim_{N\rightarrow \infty} \frac{\normb{\mu}}{N} \leq \lim_{N\rightarrow \infty}  \frac{K_o}{N} = 0.
\end{equation}
Therefore, the demand of each user $\bar{\mu}$ tends to zero  as $N$ grows.
Likewise, $\bar{\xi}\rightarrow 0$ as $N\rightarrow \infty$.
\end{proof}
}

With the previous results we can prove that the demand ratio
 $\sfrac{ \normb{\xi} }{ \normb{ \mu } }$
reaches an upper bound as the number of users increases (if the price function $p(\cdot)$ is linear).

\checkl{res:convergence_xi}{
\begin{proof}
If $v_i(x) = v_j(x) = v(x)$ for all $x\in \mathbb{R}$ and $i, j \in {\mathcal{P}}$, then all users have the same demand, that is,  $\xi_i = \bar{\xi}$ and  $\mu_i = \bar{\mu}$ for all user $i\in\mathcal{P}$.
From \cref{lemma:reducing_demand} the demand of users tends to zero as the number of users increase.
Hence, the equilibrium conditions in \cref{eq:foc_optimal,eq:foc_strategic} lead to
 \begin{equation}\label{eq:v_xi_infty}
\lim_{N\rightarrow \infty} \dot{v}_i(\bar{\xi}) = \lim_{\bar{\xi} \rightarrow 0}  \dot v (\bar{\xi}) = p( \normb{\xi} )
 \end{equation}
 and
 \begin{equation}\label{eq:v_mu_infty}
\lim_{N\rightarrow \infty} \dot{v}_i(\bar{\mu}) = \lim_{\bar{\mu}\rightarrow 0}  \dot v (\bar{\mu}) = 
\dot{C}(\normb{\mu}).
 \end{equation}
 From \cref{eq:v_xi_infty} and \cref{eq:v_mu_infty} we obtain
 \begin{equation}
 \dot{v}(0) = p( \normb{\xi} ) = p(\normb{\mu}) + \normb{\mu} \dot p( \normb{\mu} ).
 \end{equation}
 If $p(\cdot)$ is linear then 
$
 \normb{\xi} = 2 \normb{\mu}.
$
\end{proof}
}

  The next theorem shows that the price of anarchy equals zero, that is, the customer surplus in the worst case NE is arbitrarily smaller than the social optimum.

  \checkl{res:boundary_efficiency}{
\begin{proof}[Proof of \cref{res:boundary_efficiency}]
Observe that only users with positive demand 
affect
the efficiency ratio, because $U_i(0, \bs{q}_{-i}) = 0$ (see \cref{eq:profit_strategic}). 
Therefore, henceforth we assume that $\mu_i> 0$.

Now,  let us find an upper bound of the aggregate valuations at the Nash and the optimal equilibria.
On one hand, from the concavity of the valuation functions we have
$
 v_i(\mu_i) \geq v_i(0) + \mu_i \dot v_i (\mu_i) .
$
Since $\dot v_i(0) = 0 $, then  $v_i(\mu_i) \geq \mu_i \dot v_i (\mu_i)$.
Summing over all users and using 
the conditions of the optimal equilibrium in \cref{eq:foc_optimal} we obtain
 \begin{equation}\label{eq:lower_bound_v_mu}
 \sum_i v_i(\mu_i) \geq \normb{\mu} p(\normb{\mu}) + \normb{\mu}^2 \dot p( \normb{\mu} ) .
 \end{equation}
On the other hand, 
from the concavity of the valuation functions 
we obtain
$
v_i(\xi_i) \geq v_i(\mu_i) + (\xi_i - \mu_i) \dot v_i(\xi_i).
$
Likewise, summing over all agents and using the conditions of the Nash equilibrium in \cref{eq:foc_strategic} results\footnote{Observe that \cref{eq:lower_bound_v_mu} and \cref{eq:boundary_v_xi} hold with equality when the valuation functions are linear.}
\begin{equation}\label{eq:boundary_v_xi}
\sum_i v_i(\xi_i)  \geq \sum_i v_i(\mu_i) + A(\bs{\xi}, \bs{\mu}) + B(\bs{\xi}, \bs{\mu}),
\end{equation} 
where $A(\bs{\xi}, \bs{\mu}) = (\normb{\xi} - \normb{\mu})p(\normb{\xi})$ and 
$B(\bs{\xi}, \bs{\mu}) = \sum_i \xi_i (\xi_i-\mu_i)  \dot p(\normb{\xi})$.
%
%
%
%
Since $\xi_i\geq 0$ and $\normb{\xi} \geq \normb{\mu}$, 
we can obtain
the following lower bound of $B(\bs{\xi}, \bs{\mu})$
\begin{equation}\label{eq:boundary_v_xi_b}
B(\bs{\xi}, \bs{\mu}) \geq \underline{\xi} (\normb{\xi} - \normb{\mu}) = \hat{B} (\bs{\xi}, \bs{\mu})
\end{equation}

With \cref{eq:boundary_v_xi} and \cref{eq:boundary_v_xi_b} we  express the efficiency ratio as 
\begin{multline}\label{eq:ratio_a}
r(\bs{\xi}, \bs{\mu}) = 
\frac{\sum_i v_i(\xi_i) - \normb{\xi} p(\normb{\xi}) }{ \sum_i v_i(\mu_i) - \normb{\mu} p(\normb{\mu}) }
\\
\geq
\frac{ \sum_i v_i(\mu_i) + \Gamma(\bs{\xi}, \bs{\mu}) - C(\normb{\xi}) }{ \sum_i v_i(\mu_i) - C(\normb{\mu}) },
\end{multline}
where
$\Gamma(\bs{\xi}, \bs{\mu}) = A(\bs{\xi}, \bs{\mu})  +  \hat{B}(\bs{\xi}, \bs{\mu})$.

Let us represent the right hand side of \cref{eq:ratio_a} as the following function that depends on the total valuation $\sum_i v_i(\mu_i)$
%
\begin{equation}\label{wq:func_approx}
\Xi(x) = \frac{x-y}{x-z},
\end{equation}
where $x = \sum_i v_i(\mu_i)$, $y =C(\normb{\xi}) -  \Gamma(\bs{\xi}, \bs{\mu}) $ and 
$z = C(\normb{\mu})$.
%
We know that $1\geq \Xi(x)$, which implies that $y \geq z$;
therefore, the function $\Xi(x)$ is increasing with respect to $x$.

The function in \eqref{wq:func_approx} help us to find a lower bound to the efficiency ratio.
Observe that $\Xi(x) \geq \Xi(x-\epsilon)$, for $\epsilon\geq 0$.
Hence, we obtain a lower bound  
replacing $\sum_i v_i(\mu_i)$ by its lower bound in \cref{eq:lower_bound_v_mu}, resulting
\begin{equation}\label{eq:boundary_ratio}
\Xi(x) \geq \frac{ \normb{\mu}^2 \dot p(\normb{\mu}) + \hat{B}(\bs{\xi}, \bs{\mu}) - D(\bs{\xi}, \bs{\mu}) }{ \normb{\mu}^2 \dot p(\normb{\mu}) },
\end{equation}
where $D(\bs{\xi}, \bs{\mu}) = \normb{\mu} (  p(\normb{\xi}) - p(\normb{\mu}) )  $. 
Note that \cref{eq:boundary_ratio} holds with equality when the valuation functions are linear, because in that case the inequality in \cref{eq:lower_bound_v_mu} holds with equality.

Finally, if $p(\cdot)$ is linear, then the right hand side of \cref{eq:boundary_ratio} becomes
\begin{multline}\label{eq:lower_bound_eff_ratio}
\Xi(x) \geq \tilde{r} (\bs{\xi}, \bs{\mu}) 
\\
= \left( \normb{\mu}^2 + (\normb{\xi} - \normb{\mu}) (\underline{\xi} - \normb{\mu}) \right) \frac{1}{\normb{\mu}}.
\end{multline}
From \cref{lemma:reducing_demand} we know that if $\dot{C}(\cdot)$ is unbounded and all users have the same valuation, then $\xi_i \rightarrow 0$ as $N\rightarrow \infty$.
Therefore, $\underline{\xi}\rightarrow 0$, which 
 leads \cref{eq:lower_bound_eff_ratio}  to
\begin{equation}\label{eq:final_bound}
\tilde{r}(\bs{\xi}, \bs{\mu}) \rightarrow 2 - \eta, 
\end{equation}
where $\eta = \slfrac{\normb{\xi}}{\normb{\mu}}$.
From \cref{res:convergence_xi} we know that when all users have the same valuation $\eta\rightarrow 2$ as $N\rightarrow \infty$, which makes \cref{eq:final_bound} equal to zero.
Hence, the worst efficiency loss occurs when all users have the same linear valuation function, the price $p(\cdot)$ is linear, and $N\rightarrow \infty$.
%
%
\end{proof} 
}

\subsection{Properties of the Mechanism}

This subsection shows results related with the impossibility of having a mechanism that satisfies budget balance and we prove the properties of two instances of the mechanism.

Let us introduce some notation for the following results.
Let  $\theta = \sum\nolimits_{i\in\mathcal{P}} \norm{ \bs{q}_{-i} } = (N-1)\normb{q}$ and $\rho_i = \norm{ \bs{q}_{-i} } / \theta$, which satisfies $\sum_{i\in\mathcal{P}} \rho_i = 1$. 
The previous variables  allow us to rewrite the total incentives as
\begin{equation}\label{eq:sum_I}
\sum\nolimits_{i\in\mathcal{P}} I_i(\bs{q}) = \theta \left(   \sum\nolimits_{i\in\mathcal{P}} \rho_i h(\rho_i \theta) - p\left( \frac{\theta}{N-1} \right) \right).
\end{equation}


The following proof shows that it is not possible to have mechanisms with the budget balance property.

\checkl{thm:budget}{
\begin{proof}[Proof of \cref{thm:budget}]
Let us assume by contradiction that there exists some function $h(\cdot)$ such that the incentives satisfy 
$ \sum\nolimits_{i\in\mathcal{P}} I_i(\bs{q}) = 0$. 
Therefore, 
we can rewrite \cref{eq:sum_I}  as 
\begin{equation}\label{eq:condition_budget_balance}
p\left( \frac{\theta}{N-1} \right) = \sum\nolimits_{i\in \mathcal{P}} \rho_i h( \rho_i \theta ).
\end{equation}
Now, we consider two scenarios.
First, if all users have the same consumption, represented by the vector $\hat{\bs{\rho}}$ with 
$\hat{\rho}_i = \hat{\rho}_j = \frac{1}{N}$ for all $i,j\in\mathcal{P}$,
then,  \cref{eq:condition_budget_balance} becomes 
\begin{equation}\label{eq:condition_a}
p\left( \frac{\theta}{N-1} \right) =  h\left(\frac{\theta}{N}\right).
\end{equation}
Second, 
if all except
one user have positive consumption, represented by the vector $\tilde{\bs{\rho}}$ with $\tilde{\rho}_i = 0$ and $\tilde{\rho}_j = \frac{1}{N-1}$ for all $j\neq i$, then, assuming that $\lim_{\rho_j \rightarrow 0} \rho_j h(\rho_j \theta) = 0$ \cref{eq:condition_budget_balance} becomes 
\begin{equation}\label{eq:condition_b}
p\left( \frac{\theta}{N-1} \right) = h\left(\frac{\theta}{N-1}\right).
\end{equation}
Equations \cref{eq:condition_a} and \cref{eq:condition_b} are valid only when $\theta=0$ (for $N$ finite). Therefore, we find a contradiction and conclude that we cannot find a function $h(\cdot)$ that satisfies the budget balance property.
\end{proof}
}

 The next result shows the conditions to have incentive compatibility in a mechanism.
 
\checkl{res:individual_rationality_cond}{
\begin{proof}[Proof of \cref{res:individual_rationality_cond}]
Let us consider a vector $\hat {\bs{\mu}}$ equal to $\bs{\mu}$ except for its $i^{th}$ entry, which is equal to zero. 
\modification{
Hence, 
$\hat \mu_j = \mu_j$ for all $j \neq i$ and $\hat \mu_i = 0$, which leads to  $\norm{ \hat {\bs{\mu}} } = \norm{ \hat {\bs{\mu}}_{-i} } = \norm{ \bs{\mu}_{-i} } $.
}
%
From the optimality  condition in \cref{eq:foc_optimal} we know that 
\begin{multline}
W_i(\bs{\mu}) \geq W_i(\hat{\bs{\mu}})
\\
= I_i(\hat{\bs{\mu}}) = \norm{\bs{\mu}_{-i}} \left( h( \norm{\bs{\mu}_{-i}} ) - p(\norm{\bs{\mu}_{-i}}) \right).
\end{multline}
Therefore, if $ h( \norm{\bs{\mu}_{-i}} ) - p(\norm{\bs{\mu}_{-i}}) \geq 0$, then $W_i(\bs{\mu})\geq 0$.
\end{proof}
}

The next result shows an instance of the proposed mechanism that has budget deficit.

\checkl{res:h_b}{
\begin{proof}[Proof of \cref{res:h_b}]
Observe that $h(\cdot)$ in \cref{eq:h_b} is increasing, because  $p(\cdot)$ is increasing. Therefore, 
\begin{equation}\label{eq:lower_bound_h_c}
 I_i(\bs{q}) = \norm{ \bs{q}_{-i} } \left( p( \norm{ \bs{q}_{-i} } )  - p(\normb{q}) \right) \leq 0.
\end{equation} 
Hence, we conclude that $\sum_{i\in\mathcal{P}}  I_i(\bs{q}) \leq 0$.

Moreover, $h(\norm{\bs{\mu}_{-i}}) = p(\norm{\bs{\mu}_{-i}})$, satisfying the conditions of \cref{res:individual_rationality_cond}.
\end{proof}
}

The next result shows an instance of the proposed mechanism that has budget surplus.

\checkl{res:h_a}{
\begin{proof}[Proof of \cref{res:h_a}]
With \cref{eq:h_a} we can write
\begin{equation}
 \rho_i h(\rho_i \theta) =  \frac{z_i}{k \theta} p(z_i),
\end{equation}
where $z_i = \theta k \rho_i$ and $k = \frac{N}{N-1}$.
Let $t(g) = g p(g)$, which is convex (see \cref{as:2}), then 
\begin{equation}\label{eq:lower_bound_h}
\sum_{i\in \mathcal{P}} \frac{z_i}{k \theta} p(z_i) = \sum_{i\in \mathcal{P}}   \frac{N}{N k \theta} t(z_i) \geq \frac{N}{k \theta} t\left( \frac{1}{N} \sum_{i\in\mathcal{P}} z_i \right) .
\end{equation}
Here we have that $\sum_{i\in\mathcal{P}} z_i = k \theta$, therefore \cref{eq:lower_bound_h} becomes
\begin{equation}\label{eq:lower_bound_h_b}
\sum_{i\in \mathcal{P}} \frac{z_i}{k \theta} p(z_i) \geq p\left( \frac{\theta}{N-1} \right).
\end{equation}
We can use the lower bound \cref{eq:lower_bound_h_b} in \cref{eq:sum_I} to show that $\sum_{i\in\mathcal{P}} I_i(\bs{q}) \geq 0 $.

Moreover, we know from \cref{res:individual_rationality_cond} that this mechanism is individually rational since
$h( \norm{ \bs{\mu}_{-i} } ) = p(\frac{N}{N-1} \norm{ \bs{\mu}_{-i} } ) \geq p(\norm{ \bs{\mu}_{-i} })$. 
The inequality follows since $p(\cdot)$ is increasing (see \cref{as:1}).
\end{proof}
}

\modification{
\subsection{Experimental Setup}
\label{sec:experiments}

The stability of the system depends on its dynamics, i.e., on how customers update their actions in response to  price changes.
We assume that
each customer uses a load management system that finds the action (i.e., demand) that maximizes either $U_i$ or $W_i$ (if the customers receive incentives).
Here we leverage the theory of population games to define the system's dynamics that guarantee stability and convergence to 
either $\bs{\xi}$ or $\bs{\mu}$ (in the case with incentives).

In particular, we assume that each load management system has built-in some \emph{evolutionary dynamics} configured maximize the customer's profit \cite{sandholm_book, hofbauer2001nash, quijano2017role}.
The evolutionary dynamics describe how populations of agents, who participate in a game,  update their strategies to maximize a \emph{fitness function}.
We can maximize the customer's profit making the fitness function equal to the marginal profit; 
thus, the evolutionary dynamics 
resemble a distributed gradient based optimization method \cite{pantoja, mojica2015population}. 
We use the evolutionary dynamics because, thanks to the concavity of the surplus functions, 
we can guarantee that the system is asymptotically stable.

We configure the evolutionary dynamics as follows.
Each customer has some resources (e.g., energy) and decides whether to use them to perform tasks (e.g., turn on appliances). 
In other words, the customers have two strategies, whether to use or not the resources.
In this case, the benefit of using the resources equals to the marginal profit. 
On the contrary, idle  resources do not report benefits nor losses.
Thus, in the equilibrium both strategies give the same benefit, that is, the dynamics balance the benefit between both strategies. 
This also implies that in the equilibrium the marginal profit equals to zero, as required to satisfy the FOC.


We find the optimal equilibrium and the Nash equilibrium using the Population Dynamics Toolbox in \cite{toolbox}. We refer the interested reader to \cite{barreto2014incentives} for more details on the implementation.

}

\section*{References}

\bibliographystyle{elsarticle-num} 
\bibliography{references,personal}

\end{document}